\newif \ifusenix
\newif \ifacm
\newif \ifmcom
\newif \ifimwut

\usenixfalse
\acmfalse
\mcomfalse
\imwuttrue

\ifusenix

	\documentclass[letterpaper,twocolumn,10pt]{article}
    \usepackage{usenix-2020-09}
\fi
\ifacm
  \documentclass[10pt,sigconf,anonymous,authorversion]{acmart}
  \setcopyright{none}

  \acmDOI{}
    
  \acmISBN{}
    
    \acmYear{2022}
    \acmConference{IMWUT}{2022}{Virtual}

  \acmPrice{}
\fi

\ifmcom
  \documentclass{sig-alternate-10pt}
\fi

\ifimwut
\documentclass[acmlarge,screen,nonacm]{acmart}

\usepackage[english]{babel}
\usepackage{blindtext}

\AtBeginDocument{%
  \providecommand\BibTeX{{%
    \normalfont B\kern-0.5em{\scshape i\kern-0.25em b}\kern-0.8em\TeX}}}

\setcopyright{none}
\copyrightyear{none}
\acmYear{none}
\acmDOI{none}

\acmJournal{IMWUT}
\acmVolume{0}
\acmNumber{0}
\acmArticle{0}
\acmMonth{0}
\fi

\usepackage{xcolor}
\usepackage{amsfonts}
\PassOptionsToPackage{hyphens}{url}\usepackage{hyperref}
\usepackage{color,soul}
\usepackage{graphics}
\usepackage{graphicx}
\usepackage{url}
\usepackage{listings}
\usepackage{multicol}
\usepackage{multirow}
\usepackage[scaled]{helvet}
\usepackage{rotating}
\usepackage{xspace}
\usepackage{algorithm}
\usepackage{comment}
\usepackage{enumitem}
\usepackage{amsmath}
\usepackage{mathrsfs}
\usepackage{cancel}
\usepackage{cleveref}
\usepackage[font=small,labelfont=bf]{caption}
\usepackage[skip=0pt]{subcaption}
\newcommand{\name}{R-fiducial\xspace}
\newcommand{\hlcolor}{black}

\begin{document}


\makeatletter
\def\runningfoot{\def\@runningfoot{}}
\def\firstfoot{\def\@firstfoot{}}
\makeatother

\settopmatter{printacmref=false} 
\renewcommand\footnotetextcopyrightpermission[1]{} 
\pagestyle{plain} 


\title{\name: Reliable and Scalable Radar Fiducials for Smart mmwave Sensing}

\author{Kshitiz Bansal}
\authornote{Both authors contributed equally to this research.}
\email{ksbansal@eng.ucsd.edu}
\author{Manideep Dunna}
\authornotemark[1]
\email{mdunna@eng.ucsd.edu}
\affiliation{%
  \institution{University of California, San Diego}
}

\author{Sanjeev Anthia Ganesh}
\affiliation{%
  \institution{University of California, San Diego}
  \city{New York}
  \country{USA}}
\email{santhiag@ucsd.edu}

\author{Eamon Patamasing}
\affiliation{%
  \institution{University of California, San Diego}
}
\email{eamon@ucsd.edu}

\author{Dinesh Bharadia}
\affiliation{%
  \institution{University of California, San Diego}
  }
\email{dineshb@eng.ucsd.edu}

\makeatletter
\def\@maketitle{\newpage
 \null
 \setbox\@acmtitlebox\vbox{%
  \begin{center}
    {\ttlfnt \@title\par}       
{\subttlfnt \the\subtitletext\par}\vskip 1.25em
    {\baselineskip 16pt\aufnt   
     \begin{tabular}[t]{c}\@author
     \end{tabular}\par}
  \end{center}}
 \dimen0=\ht\@acmtitlebox
 \unvbox\@acmtitlebox
 \ifdim\dimen0<0.0pt\relax\vskip-\dimen0\fi}
\makeatother


\begin{abstract}
    Millimeter wave sensing has recently attracted a lot of attention given its environmental robust nature. In situations where visual sensors like cameras fail to perform, mmwave radars can be used to achieve reliable performance. However, because of the poor scattering performance and lack of texture in millimeter waves, radars can not be used in several situations that require precise identification of objects. In this paper, we take insight from camera fiducials which are very easily identifiable by camera and present \name tags which smartly augment the current infrastructure to enable a myriad of applications with mmwave radars. \name acts as a fiducials for mmwave sensing, similar to camera fiducials and can be reliably identified by a mmwave radar. We identify a precise list of requirements that a millimeter wave fiducial has to follow and show how \name achieves all of them. \name uses a novel spread-spectrum modulation technique that provides low latency with high reliability. Our evaluations show that \name can be reliably detected with a 100\% detection rate upto 25m and upto 120 degrees field of view with a latency of the order of milliseconds. We also conduct experiments and case studies in adverse and low visibility conditions to showcase the applicability of \name in wide range of applications. 
    
   
\end{abstract}

\begin{teaserfigure}
    \centering
     \includegraphics[width=\linewidth]{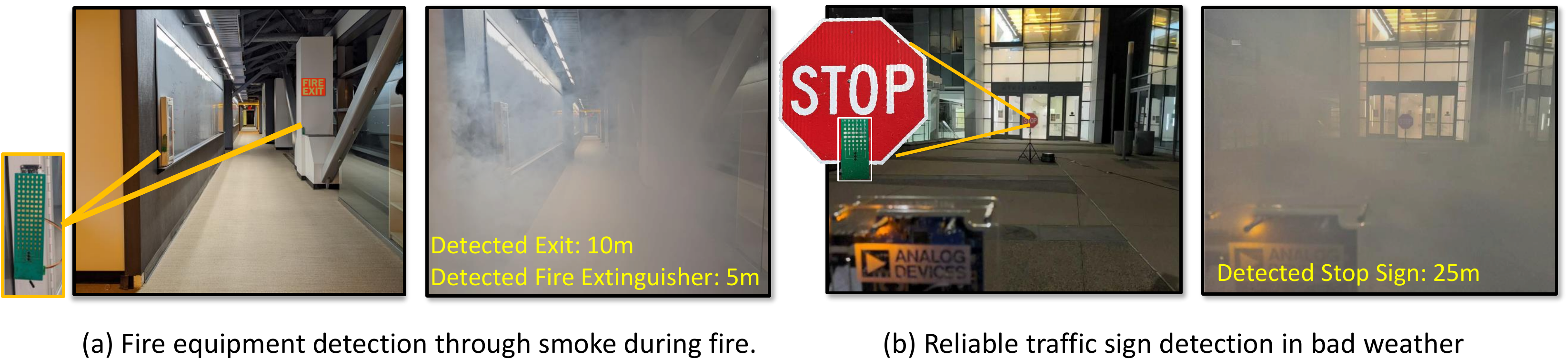}
    \caption{\name enhances the ability of mmwave radars by augmenting the current infrastructure with specialised hardware. (a) Shows a corridor filled with smoke during fire where camera fails to detect fire equipment and exit. (b) Shows a stop sign in foggy weather. In both the cases the presence of \name tags enables the radar to reliably identify and localizes fire extinguisher and exit in (a) and stop sign in (b).}
    \label{fig:first-fig}
\end{teaserfigure}
\maketitle

\section{Introduction}
\label{sec:intro}

Millimeter wave (mmwave) radar sensing has made significant advancements with applications in both indoor and outdoor areas because they are environmentally agnostic compared to their visual counterparts. For example, starting from indoor applications like people counting~\cite{weiss2020improved}, building security~\cite{stanko2008active}, smart home~\cite{dong2020secure, liu2020real} and health monitoring~\cite{alizadeh2019remote, zhao2020heart} to outdoor systems like blind spot detection~\cite{liu2017blind}, adaptive cruise control~\cite{kuroda1998adaptive}, traffic monitoring~\cite{jin2020mmwave} and drone perception~\cite{dogru2020pursuing}, mmwave radars pervades almost all aspects of our everyday life. 


One would imagine that given the advantages of mmwave radars in visually adverse conditions, we should use them in all the applications where cameras are used. Applications that require identifying objects in an environmental-independent way like stop signs in poor weather conditions~\cite{fang2003road} or fire extinguishers in low visibility conditions like smoke and fire~\cite{starr2014evaluation}, would heavily benefit from using mmwave radars. These radars also provide a longer range than cameras ~\cite{meyer2019automotive}. However, with the current mmwave radar systems it becomes very challenging to achieve those applications which involve precisely identifying objects. The reason behind this limitation arises from two fundamentally different physical properties of mmwaves, compared to the nanometer wavelength at which camera operates -- poor scattering performance and lack of an identifying signature like color. For example, when a mmwave radar wants to read a small roadside stop sign, poor scattering performance will limit the power reflected back towards the radar. Even if some signal is received back by the radar, mmwave lacks the rich color information unlike camera, which can help identify a particular road sign. As these limitations of mmwave radars are physical limitations, it is not possible to overcome them by any kind of processing alone.

Given that the lack of color and poor scattering of mmwaves are the root causes of their limitations in identifying objects, creating a smart infrastructure that is guaranteed to be reliably identified by radars, would solve the problem. This is similar to the camera domain, where fiducials like aruco markers~\cite{marut2019aruco} provide an anchor point that is very easily identifiable by the camera, regardless of the background. To design such radar fiducials that can be used ubiquitously in both indoor and outdoor applications, the following requirements have to be met:

\textbf{A) Large field of view and range:}
\label{subsec:retro}
The primary requirement of a radar fiducial is that the radar should be able to identify it, regardless of the tag's orientation. For example, if the radar fiducial is used to mark a fire extinguisher in a corridor that is 2m wide, it should have more than 120 degrees field-of-view (FOV) to be identifiable by the mmwave radar that is 5m away. This implies since both transmitters and receivers are at the same place in a mmwave radar, the signal reflected from the radar fiducial should send a reasonable amount of power back in the same direction as the incident signal for the radar to identify the radar fiducial and achieve a wide FOV. Additionally, it is important that these fiducials are identifiable from long distances specially in the case of outdoor applications, thus needing large FOV and range for these radar fiducials.

\textbf{B) Reliable identification:}
\label{subec:det_iden}
In a real world deployment, radar would also receive reflections back from all the static and dynamic objects present in the scene. All these signals would act as clutter for the signal reflecting back from our fiducial. This further imposes that these fiducials be designed such that when scanned by a radar they:(i) are uniquely identifiable by the radars, (ii) do not raise false alarms by confusing them with other objects (false positives), and (iii) never be missed of their presence (false negatives). For example, a fiducial present on a fire extinguisher should identify itself as a fire extinguisher while the one on exit sign should identify itself as an exit sign. Radars should always report their existence in the FOV and no other objects should be falsely identified as these fiducials (fire extinguishers and exit signs). Hence identification in a reliable way forms the second core requirement of radar fiducials.


\textbf{C) Low Latency:} 
Some of the target applications of radar fiducial are life-critical with the strict conditions of low-latency operation to prevent calamities. For example, a car moving at 30 mph requires at least 25m to stop after the driver is alerted, \cite{braking_distance} which means that the latency of detecting a stop sign needs to be of the order of milliseconds to ensure a safe stopping distance. Furthermore, the required latency of detection is directly proportional to the distance from the stop sign, i.e., if the car is closer to the stop sign, it should detect at a much lower latency. Thus radar fiducials need to have a latency of the orders of milliseconds. \textcolor{\hlcolor}{Also, the sensor frame rates have a major impact on stopping distances. A higher frame rate would help reduce the stopping distance \cite{mody2016adas}, making driving much safer. Generally, most automotive sensors operate at around 20-30 frames per second which means that the tag should be readable within the millisecond time frame (30 fps corresponds to 33ms frame time).}

\textbf{D) Compatibility with existing radars:} 
To ensure wide-applicability of radar fiducials, they should be compatible with existing radars.
Most of the existing radars operate in two different modes~\cite{dham2017programming} -- \textit{digital beamforming} mode and \textit{analog beamforming} mode. In digital mode, the signal is sent uniformly across the space in each measurement, while in analog mode, different angles are scanned in different measurements. So, firstly, these radar fiducials should not require any hardware change on the current widely used mmwave radars. Secondly, both digital and analog beamforming modes of operation in the current radars must be supported.

\textbf{E) Accurate Localization:}
Finally, the target applications of radar fiducials such as searching for fire-extinguisher in case of fire and stop sign detection would also require accurate localization of the fiducials along with reliable identification. This means that while the radar fiducial creates a unique identity of itself for the radar, it should also make sure that the radar can locate it with the traditional radar processing. Hence, another requirement for radar fiducial is to be accurately localized by the radars.



Achieving all the above requirements in a single design is a challenging task. Past work~\cite{soltanaghaei2021millimetro,nolan2021ros,mazaheri2021mmtag} that focuses on developing smart mmwave infrastructure falls short of satisfying one or more of the above requirements. Millimetro\cite{soltanaghaei2021millimetro} provides a good range but the identification scheme does not guarantee reliability (requirement B), low-latency (requirement C) and compatibility with all radars (requirement D). RoS~\cite{nolan2021ros} develops specialized hardware which changes the polarization of the signal and thus requires specialized radar design to receive different polarizations, violating requirement D. RoS~\cite{nolan2021ros} and mmtags~\cite{mazaheri2021mmtag} are also limited to short ranges (requirement A). A design that can truly act as a radar fiducial is still non-existent.

In this work, we present \name tags that act as radar fiducials and achieve all the above requirements in a single design. We describe, how \name's careful design choices and novel detection algorithms holistically address the requirements of range, FOV, reliable identification, localization and low-latency, while keeping them compatible with all types of mmwave radars. \name tags low-power and low-cost design allows them to be scalable and easily deployable.



The first challenge for \name is to design a tag that can achieve a large FOV and range (requirement A). To achieve this, the tag should reflect enough power back towards the radar regardless of the angle of incidence. However, poor scattering characteristics of mmwaves reduce the power that reflects back towards the radar, hence limiting the FOV to near-perpendicular incidence. Using multiple antennas on the tag will improve the gain but will not cause the signal to reflect back towards the radar. Thus, to achieve wide horizontal FOV \name uses the van-atta array concept~\cite{sharp1960van,vanatta_patent} because of its ability to \textit{retro-reflect} the signal without active elements i.e. reflecting the signal in the same direction. While there are existing works that use these van-atta array tag designs~\cite{soltanaghaei2021millimetro,nolan2021ros,mazaheri2021mmtag}, \name presents a new framework for designing a van-atta architecture by defining two major variables: the number of van-atta pairs and the number of patches per antenna. We analyze the practical implications of choosing different architectures in this design space defined by these two variables while taking into account the requirements of achieving a longer range (getting a higher gain) and supporting wider vertical FOV. 

Next, the \name tags should be uniquely identifiable by mmwave radars in real-world environments (requirement B). This means that these tags should create a signature for each tag that is different from the background objects and unique for each tag. Moreover, it is extremely important to have the reliability of operation (low false positive and false negative rate). To create a unique signature, a simple solution is to modulate the signal reflected from the tag, similar to RFIDs. This can be achieved by adding a switch in the van-atta design and using an ON-OFF scheme to modulate the signal that gets reflected from the tag as done in~\cite{soltanaghaei2021millimetro}. However, using this scheme poses serious concerns for the reliable identification of the tag. A simple periodic ON-OFF switching scheme creates signatures in the doppler domain that can easily be confused by the doppler of other objects present in the environment. This leads to either false positives (wrong detections) or false negatives (mis-detections). To solve this problem, \name designs a novel modulation scheme that uses \textit{spread-spectrum codes} and ensures reliable detection and identification of the tags(requirement B). Moreover, \name's modulation scheme allows for sensing the tag within a single radar measurement which is called chirp, while supporting multiple tags operation.

Another favorable outcome of \name's single-chirp modulation is that it also provides low-latency (requirement D) and compatibility (requirements E) with all radar modes in a single design. Even an analog beamforming radar can read \name with just a single chirp sent towards the tag. Existing modulation schemes~\cite{soltanaghaei2021millimetro} use multi-chirp modulation that puts a lower limit on the achievable latency and renders them incompatible with analog-beamforming radars. On the other hand, \name's modulation scheme allows it to be identified within a single chirp duration, providing the least possible latency, compatibility with all radar types and maximum reliability, a feature that can not be attained by existing art.


For ubiquitous deployment of \name tags, we develop algorithms to detect, identify and localize these tags (requirement E). As mentioned before, one of the most vital things is to have minimal requirements on the radar reading hardware. To showcase our system, we develop algorithms for mmwave radars, which enables them to uniquely isolate and identify multiple tags present in a complex real-world scene, without any changes to the hardware. To perform decoding, \name needs to identify the spread spectrum code from the back-scattered signal. Not having apriori knowledge about the distance between the tag and the radar poses a unique challenge for the decoder. To solve this, we provide an algorithm that jointly solves for the distance as well as the code present on the tag. Unlike prior works, \name's code design and algorithm are agnostic to very high doppler(up to 150 kmph) caused by the fast-moving vehicles and can localize and detect the tags with a median localization error of 0.2m.


We prototype the tag antenna design with low-power RF switches that can last for more than a year on CR2477 coin cell\cite{cr2477_datasheet} while modulating the tag with a code continuously. Our tag provides a wide horizontal FOV of more than 120 degrees. We manufacture our design and evaluate it comprehensively in indoor and realistic outdoor driving settings using commercial off-the-shelf (COTS) 24 GHz radar with 2 TX and 4 RX antennas. With our prototype, we showcase an application of traffic infrastructure augmentation and an indoor application of object identification (figure~\ref{fig:first-fig}). In outdoor application, the radar is deployed on a car, and the performance is evaluated at different speeds, while in indoor application multiple tags are deployed in a multi-path rich environment. Our evaluations show that \name can be detected with \textbf{100\%} detection rate up to \textbf{25m} of range. \textcolor{\hlcolor}{Moreover, \name can be localized within a median angle error of \textbf{6} degrees and a median ranging error of \textbf{0.1}m, which is well within the requirement of the target applications. We also evaluate the reliability of \name when multiple active tags are present in the environment and achieve an average error of less than 30cm.} A video demonstration of \name could be found at \textit{https://streamable.com/7ax59s} 



\begin{figure}[t!]
    \centering
    \includegraphics[width=\linewidth]{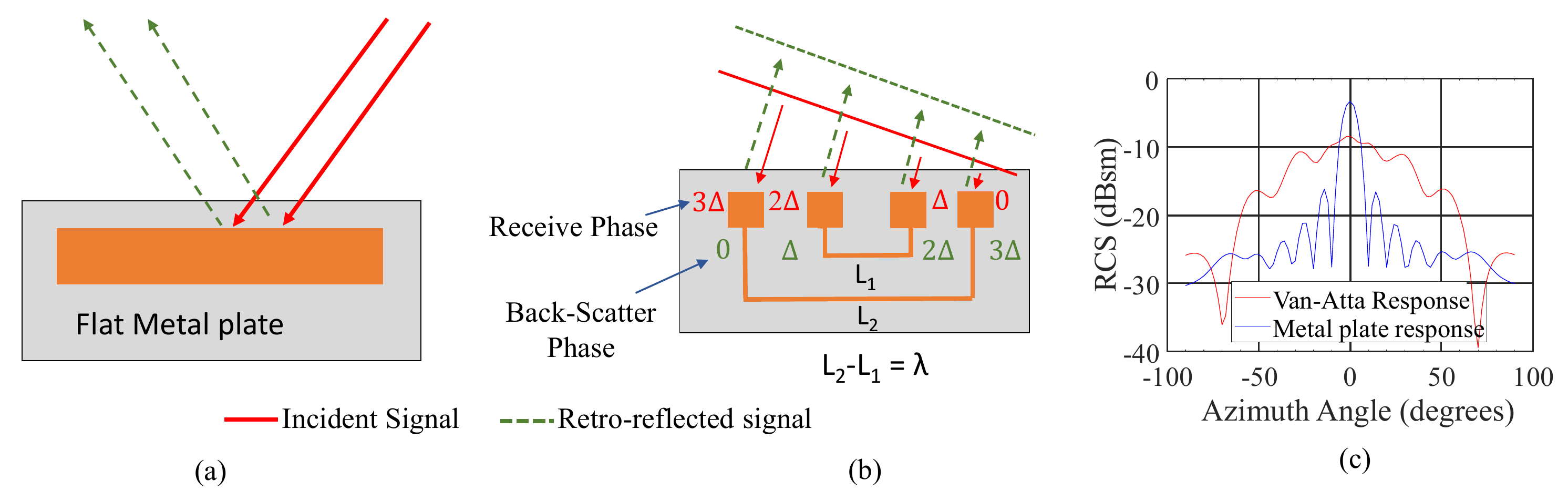}
    \caption{{a) Flat metal plate reflecting incident signals b)Retro reflective property of Van atta arrays c) The Monostatic response of the flat metal plate versus the Van-atta array}}\label{fig:van_antenna_pairs}
\end{figure}

\section{Design of \name}
Based on the requirements described in section \ref{sec:intro}, we present the design of \name. Specifically, in this section, we show how \name fulfills the requirement of large FOV along with a long range. We start by defining the design space of making a tag that can act as a mmwave fiducial and provide a framework to optimize the tag design within that space based on the requirements.

\subsection{Van-atta architecture and RCS primer:} To meet requirement A of the wide field of view, our design is inspired by the van-atta array concept~\cite{sharp1960van,vanatta_patent} discovered in early 1960, wherein the incident RF signal is reflected back in the same direction. The key condition here, that achieves retroreflectivity, is to ensure the paths connecting the antennas have a length difference equal to an integer multiple of the wavelength. To see how this condition enables retroreflectivity, consider a plane wave incident at an angle. Each adjacent antenna incurs an additional phase delay with respect to the previous antenna. The paths that connect the antenna pairs all induce the same phase because of an integer multiple path length difference. This means that the pair of antennas that are connected to each other exchange their phases. In this way, each antenna pair achieves the same wave-front in the direction of incidence thereby achieving retro-reflectivity. Figure~\ref{fig:van_antenna_pairs}b shows how a van-atta arrangement achieves retro-reflectivity. 

The reflected power is measured in terms of RCS (Radar Cross Section) which is a proxy for the strength of reflections from an object. RCS is a function of both incident and reflection angles. As the automotive radar has co-located Tx and RX, we are interested in mono-static RCS (Tx and Rx are in the same direction) values at different angles in the azimuthal plane. Figure\ref{fig:van_antenna_pairs}c plots the monostatic RCS of a Van-Atta array and compares its RCS with a flat metal plate. The flat metal plate(figure\ref{fig:van_antenna_pairs}a) reflects the incident signal back to the Tx only when the Tx is present along the normal direction(i.e at 0-degree incidence) to the metal plate as indicated by the pointed RCS curve for the flat metal plate. This behavior is due to the specular reflective property of flat metal sheets. On the other hand, the Van-Atta array reflects the incident signal back to the Tx over a wide range of incidence angles.  

    

\begin{figure}[t!]
    \centering
    \begin{minipage}{0.28\textwidth}
     \includegraphics[width=\linewidth]{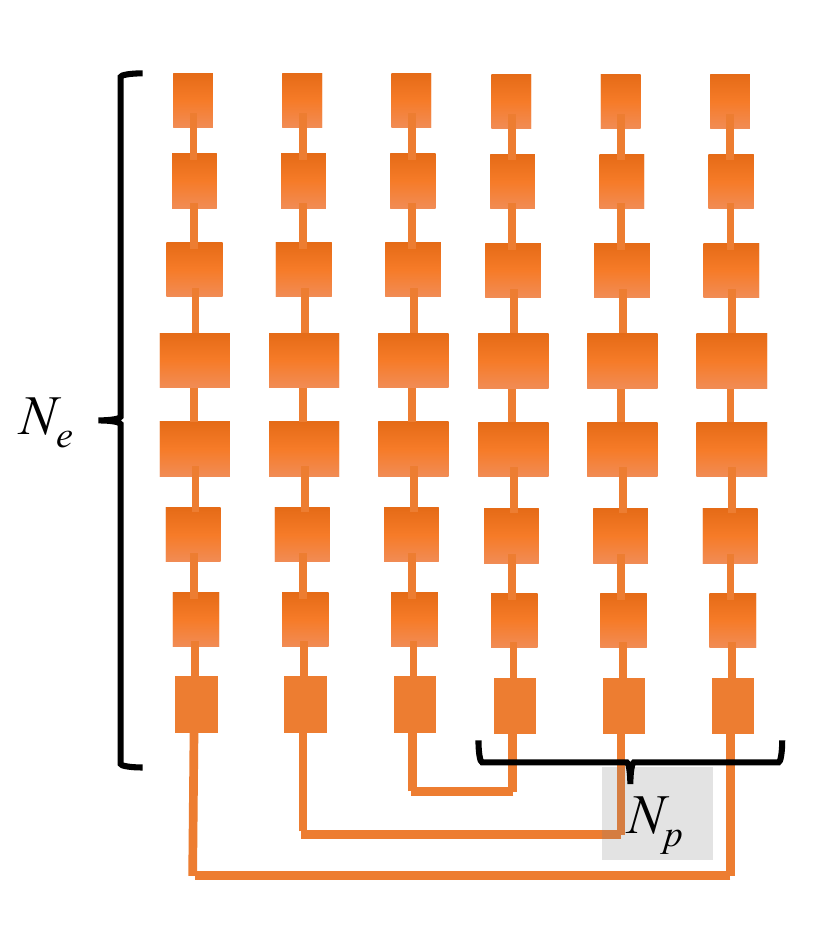}
     \subcaption{}
    \end{minipage}
    \hfill
    \begin{minipage}{0.37\textwidth}
    \centering
     \includegraphics[width=\linewidth]{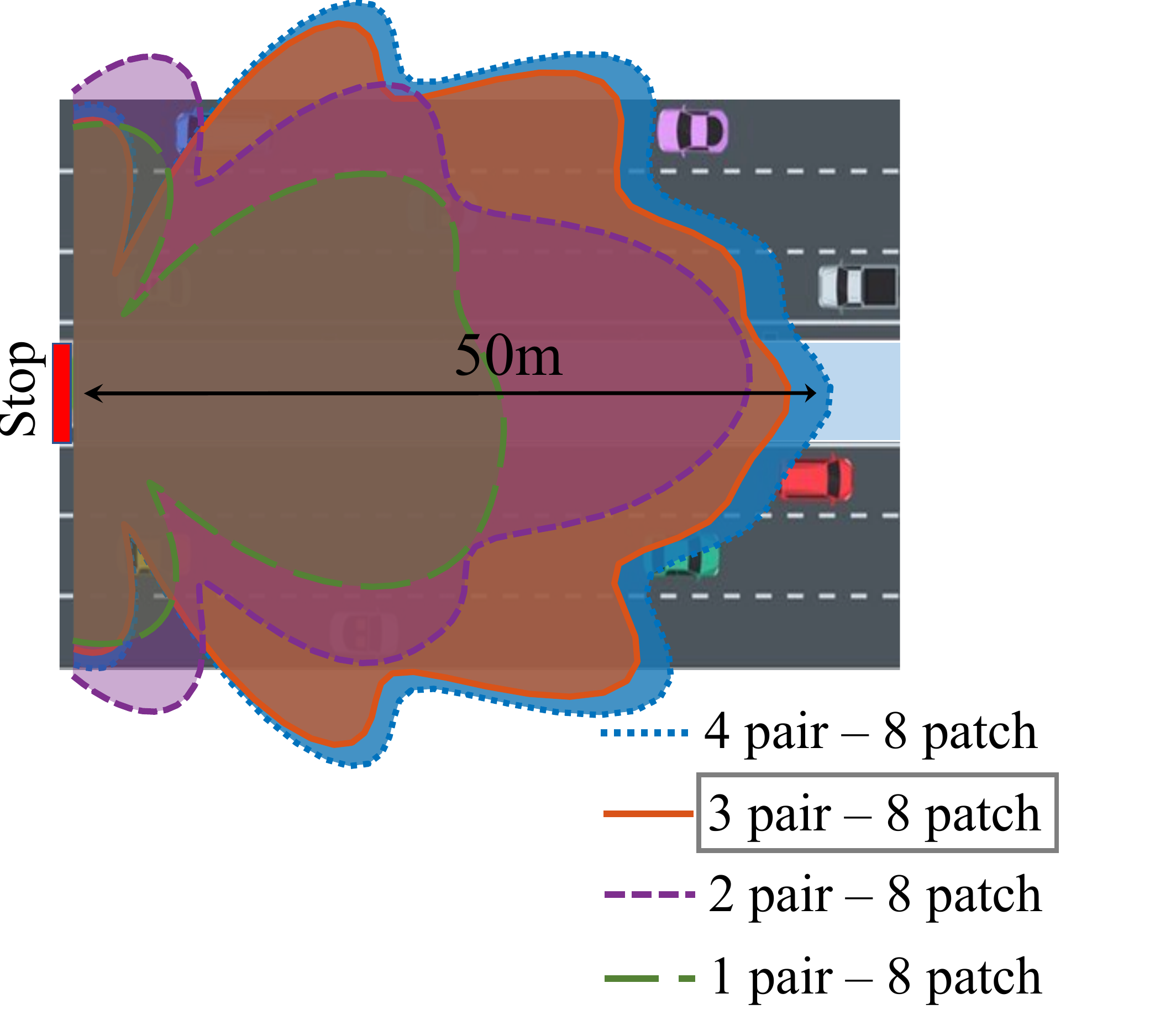}
     \subcaption{}
    \end{minipage}
    \begin{minipage}{0.31\textwidth}
    \centering
     \includegraphics[width=\linewidth]{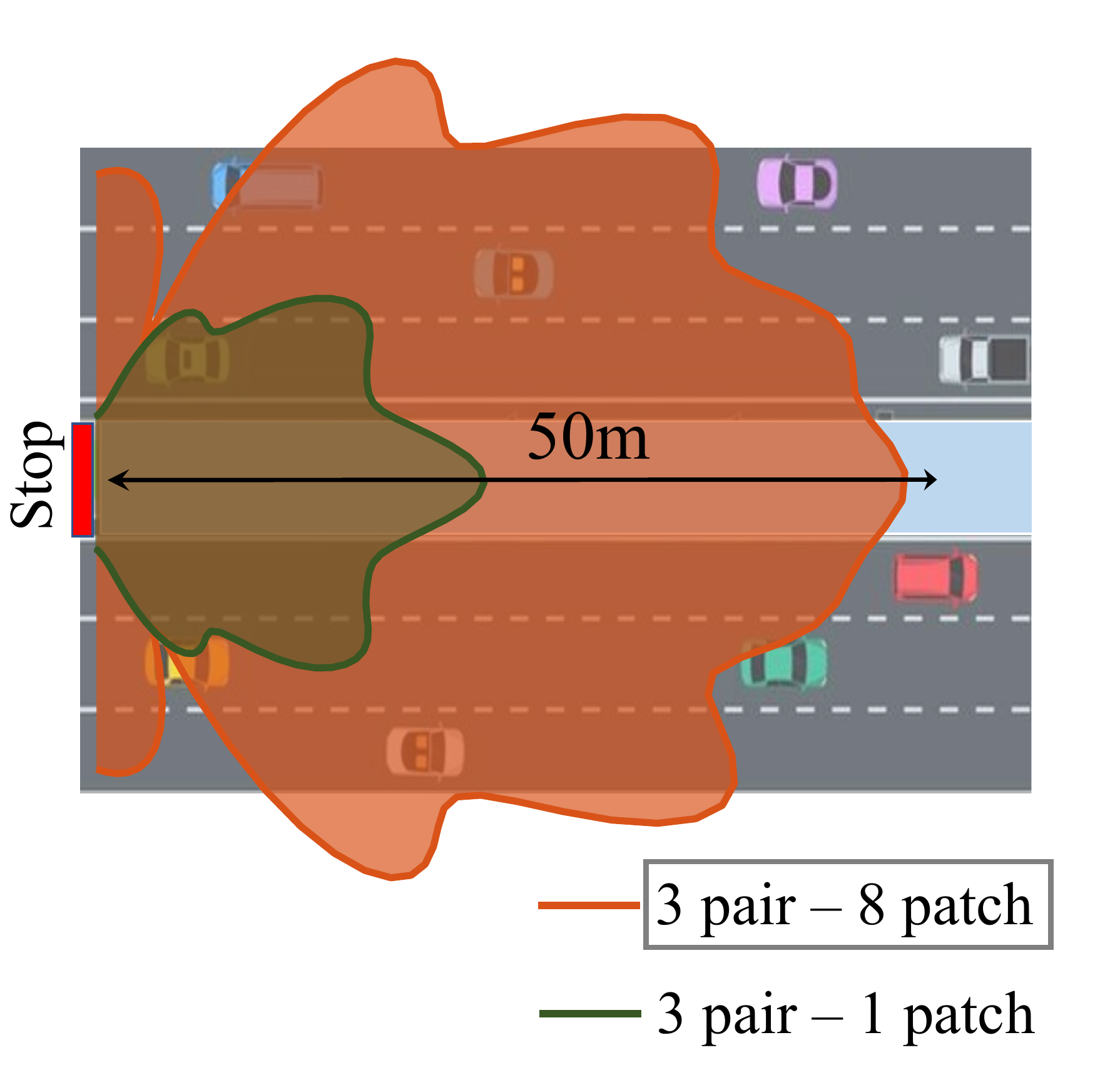}
     \subcaption{}
    \end{minipage}
    \vspace{3pt}
    \caption{{a) Designed 3 pair 8 element series-fed patch antenna based van atta array b) Coverage area comparison for 1,2,3,4 pair of antennas with 8 patch elements on each antenna. Beyond 3 pairs in the array, there is no significant improvement in range. c) Coverage area comparison with 8 patches per antenna and 1 patch per antenna. 8 patches show a significant improvement  in the range due to the vertical beamforming gain.}}\label{fig:coverage}
\end{figure}



\subsection{Choosing a van-atta architecture}
A van-atta array design can achieve the requirement of wide horizontal FOV, but for any practical use case, we also need to ensure a reasonable range of detection and vertical FOV. These 3 requirements together form the basis for designing a van-atta array architecture for any mmwave fiducial application. Next, we will describe the effect of choosing different van-atta architectures on the above three requirements.

The tags we are designing consist of series-fed patch antenna arrays\cite{series_fed}, which are easily printable on circuit boards. Now, for a van-atta array architecture, the two main design variables are the number of connected patch elements per antenna ($N_e$) and the number of antenna pairs ($N_{p}$)  (Figure~\ref{fig:van_antenna_pairs}b). Here we will provide a framework to choose $N_e$ and $N_p$ while designing the tags. First, we will start with analyzing the implications of the number of patch elements in each antenna. Having more patch elements per antenna (higher $N_e$) would increase the beamforming gain of the antenna in the vertical plane thereby increasing the RCS of the tag. $N_e$ patches in series, constructively combine the reflected signal and increase the RCS by a factor of 20log($N_e$) dB when compared to an antenna made with a single patch. Figure~\ref{fig:coverage}c shows the improved coverage of the tag when we move from $N_e$=1 to $N_e$=8 for a 3-pair antenna array. Now, using a very large $N_e$ is a tempting choice as it provides a higher gain and thus a longer detection range. However, a large $N_e$ would also mean a narrower vertical beamwidth(smaller Field of View), which would impact the radar elevation from the ground. So, $N_e$ has to be chosen such that the vertical FoV of the tag fulfills the application requirements. 

\begin{figure}[t!]
    \centering
    \includegraphics[width=0.8\linewidth]{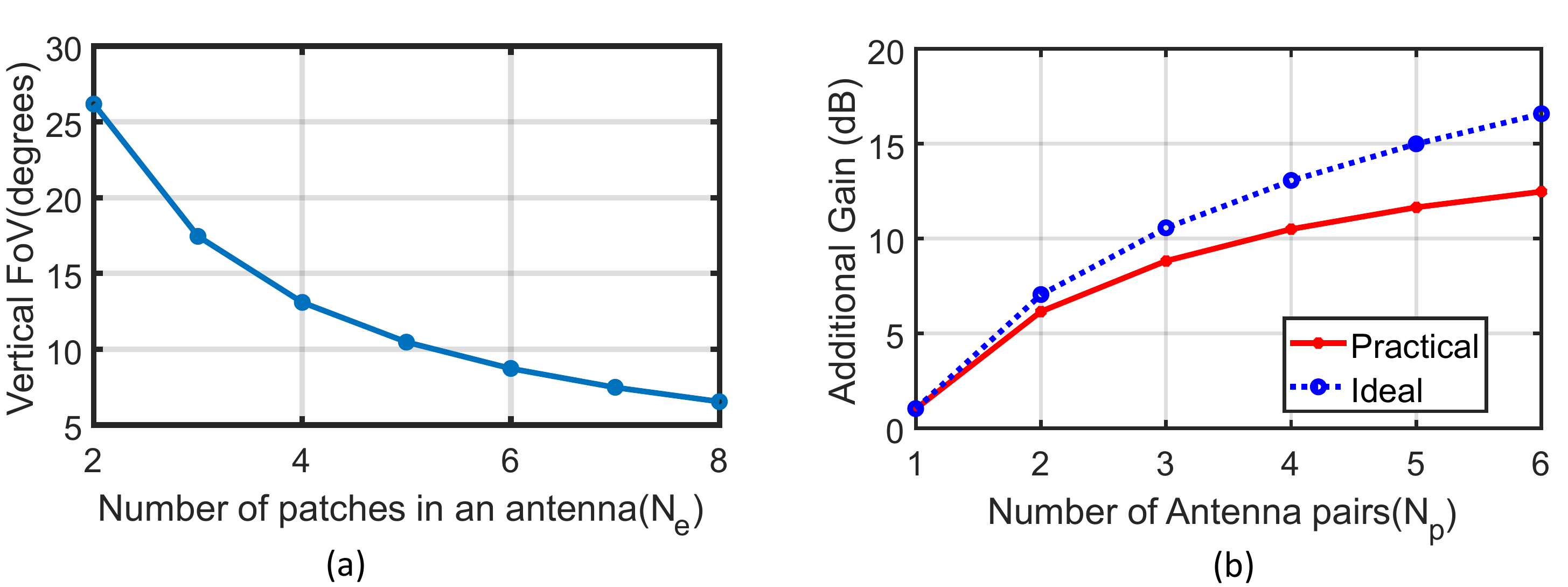}
    \caption{{a)Vertical Field of View(FoV) of the tag vs number of patches in an antenna b) Comparison of Gain for different number of antenna pairs in the ideal case with the Lossy case.}}
    \label{fig:Np_Ne_tradeoff}
\end{figure}

To more concretely understand the effect of $N_e$, let us take an example of an outdoor application of traffic sign detection. We mount these tags on a road-side sign and identify these tags using automotive mmwave radars. In such a scenario, we want the tags to be detected by radars mounted on any vehicle at any height. Due to wide variance in vehicle height, the tags must be designed such that the radars placed at different heights on the vehicle must be present in the vertical FoV of the tag. For example, considering a maximal height of 4m\cite{Truck_height}, 4m elevation should be in the FoV of the tag when the vehicle is around 45m away from the tag, based on DMV requirements\cite{signal_200ft}. To fulfill this requirement, FoV of the tag should be close to $tan^{-1}(4/45) \approx 5.07$degree. The 3 dB beamwidth\cite{balanis2015antenna} of the antenna is given as $\frac{0.886\lambda}{2N_ed}$ where d is the separation between patch elements in an antenna and $\lambda$ is the wavelength. Figure\ref{fig:Np_Ne_tradeoff} plots the beamwidth for the different number of patches in the antenna. If we choose 8 patch elements with half-wavelength separation between neighboring elements, we fulfill this requirement. Similarly, for an indoor use case of identifying fire safety equipment in an adverse situation, we can choose a narrower vertical FOV as most equipment are typically located at a low elevation from the ground for easy accessibility. In this case, one can design tags to have double the patch elements per antenna to detect them more reliably. \textcolor{\hlcolor}{Note that the patch sizes in a single series-fed patch antenna element are chosen to be of varying sizes. Although the most common series fed patch antennas have identical patch elements in an antenna unit to provide gain, they create side-lobes in unwanted directions. This is similar to the problem of having high side-lobe levels in the frequency response of a rectangular window(identical coefficients). The solution to this problem is to design windows with weighted coefficients (Eg: Hanning window, Chebyshev window) to reduce the side-lobe power level. Following this analogy, we vary the antenna sizes in a series fed antenna according to the Dolph Chebyshev window to achieve low side-lobe radiation and put most of the radiated energy in the main lobe of the antenna pattern. We have chosen the Dolph-chebyshev window because of its constant side-lobe level property. For this, we have tapered the antenna sizes following the Dolph-Chebyshev pattern synthesis method\cite{dolph1946current,DC_series_fed}}. 

Recall that we have another design parameter, the number of antenna pairs $N_p$. A higher number of pairs($N_p$) improve the beamforming gain of the array in the horizontal plane, retro-reflecting more signal power and resulting in a higher RCS. Although the number of pairs theoretically increases the RCS, there is a practical limitation on the number of antenna pairs due to associated trace losses on the tag.




\textbf{Effect of Trace losses:} Theoretically, $N_p$ pairs of antennas should give an additional $20log(N_p)$ dB gain compared to a single pair. But, in practice, the transmission lines connecting the pairs of antennas have a loss factor due to the lossy PCB substrate and increased copper resistance at millimeter-wave frequencies. For example, transmission line(TL) trace lengths in a Np pair array are of lengths $0,\lambda,2\lambda,..,(N_p-1)\lambda$ where $\lambda$ is the wavelength at 24GHz. The trace losses are proportional to the length of the line joining the antennas and the trace loss increases as we add an additional pair of antennas. The trace loss is simulated using the loss tangent of RO4003C substrate that we used in our PCBs. Figure \ref{fig:Np_Ne_tradeoff}b shows the additional gain we would obtain as we increase the number of antenna pairs and compares it with the realized array gain by including the substrate losses. We see that the difference between the Ideal gain and the theoretical gain starts to increase implying losses in the substrate tend to dominate for a higher number of pairs. Beyond 3 pairs, the increase in the gain is marginal. We perform simulations in HFSS(an ElectroMagnetic field simulator) to validate our design. We obtained the RCS for different antenna pairs and plotted the corresponding range in Figure \ref{fig:coverage}b. We found that when we have more than 3 pairs, the additional range we achieve is marginal. So, we used 3 pairs of antennas in our design. We choose our operating point at $N_p$ = 3 and $N_e$ = 8 (figure~\ref{fig:coverage}a). With our architecture, we achieve -8 dBsm of peak RCS providing us a range of 45 meters (we use the standard noise and gain parameters as provided by our radar's datasheet. Noise Figure = 10dB, Transmit power = 8 dBm, antenna gain = 9dB), which is sufficient for our application. Note, that depending on some specific application, some other operating points can also be chosen based on the analysis framework provided above. 

\section{Detecting the presence of tag: Modulation}
\label{sec:modulation}
In the previous section, we have explained the design of our tags that can retroreflect the incident signal back to the radar meeting the requirement of a large FOV. However, in a realistic environment, the radar would be receiving reflections from numerous other objects which would act as clutter in identifying the existence of the tag's reflections. Thus we need to enable \name to reliably and accurately identify the tag in the presence of other objects in the environment. Furthermore, \name should also enable scalability to multiple tags as there can be multiple objects which need to be identified. In this section we would provide the details of our solution where we provide a unique signature to the \name's reflected signal and use it for identification. With our modulation scheme, we achieve the requirements of reliable identification, low latency and compatibility. 

\subsection{Radar Primer}
\label{sec:radar_primer}
Most commercial mmwave radars, use frequency-modulated carrier waveform (FMCW or chirp signal) to find the location of the objects and the speed at which they move in the environment. An FMCW radar works by transmitting a chirp signal that sweeps a bandwidth B within a T time duration. The radar processes the reflections coming off the objects to infer the locations of the objects and the speed at which the objects in the environment move relative to the radar. For example, a tag/object that is present at a distance $d$ from the radar, the transmitted chirp signal reaches the tag/object, reflects, and reaches back to the radar traveling a round trip distance $2d$. The round trip travel delays the received chirp by $\tau = \frac{2d}{c}$ where $c$ is the speed of light. The round trip delay causes an instantaneous frequency difference between the transmitted and received chirp equal to $\Delta = \frac{B}{T}\tau$. The radar de-chirps the received signal by conjugating it with the transmitted chirp and then the Analog to Digital converter (ADC) at the radar digitizes the de-chirped signal and generates the radar samples. The instantaneous frequency($\Delta$) is estimated by taking the FFT of the ADC samples in a chirp(Range FFT). The Range bin at which the peak appears in the FFT corresponds to the frequency difference $\Delta$ which is then used to calculate the separation between the object in the environment and the radar. 

When the object is moving relative to the radar, the received chirps get shifted by a frequency proportional to the relative velocity of the object also known as the doppler effect. The frequency shift is given by $\frac{v}{c}f_c$ where $f_c$ is the center frequency of the chirp and $v$ is the relative speed of the object. This doppler frequency is manifested as a changing phase in the received signal between two consecutive chirps. We take an FFT along multiple chirps (Range-Doppler FFT) to estimate the doppler frequency shift. The Doppler bin where the peak in the Doppler FFT appears gives the doppler shift which is then used to estimate the relative speed of the object. \textcolor{\hlcolor}{The doppler bandwidth of a radar depends on the time between the start of two consecutive chirps. For example a radar with a 600$\mu$s time duration between the start of two consecutive chirps, we can have a maximum detectable doppler frequency of only 1.66 kHz.}

\subsubsection{Resolution and Error in the measurement}
\textcolor{\hlcolor}{Using an FMCW based architecture imposes a limit on the resolution that a radar system can achieve. The resolution determines the minimum separation necessary in order to resolve two reflections. For distance measurement, this resolution limit is dependent on the total bandwidth of the chirp waveform and is given by $\frac{c}{2B}$. For doppler estimation, the resolution depends on the total number of chirps that are coherently processed in a frame. For angle estimation, the resolution depends on the total size of the aperture (in multi antenna receive system it is sometimes equivalent to the total number of receive antennas). The error in the distance measurement depends on how fine is the bin-width of fft, i.e., number of fft points and the signal to noise ratio(SNR) of the received signal. Thus, by increasing the number of fft points(by zero padding the radar samples) one can decrease the error in the distance estimation~\cite{bhutani2019role}. Note that the error would still be affected by the received signal strength which decreases at longer distances and consequently estimation error increases at farther distances.}

\begin{figure}[t!]
    \centering
    \includegraphics[width=\linewidth]{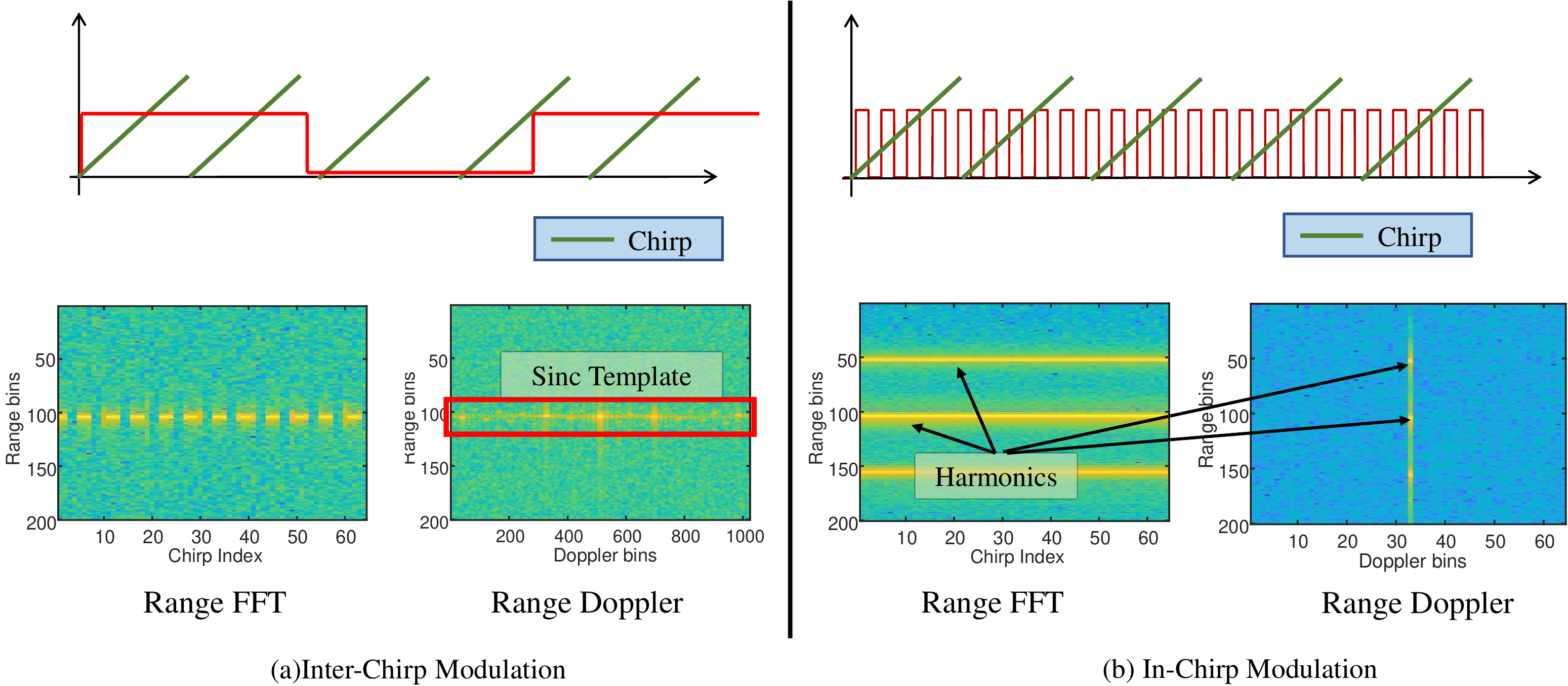}
    \caption{{Figure showing a)Inter-chirp modulation where the modulation time period is much smaller than that of the chirp and encodes the tag modulation over multiple chirps. Range Doppler plot shows the sinc template formed in the Doppler domain due to the ON-OFF modulation. b) In-chirp modulation where the modulation period is much smaller than a chirp duration and encodes the tag modulation in a single chirp. Range FFT plot shows the harmonics created in every chirp.}}
    \label{fig:in_vs_inter_chirp}
\end{figure}

In a typical indoor/outdoor setup, the radar signals will be reflected off many objects and arrive back at the radar. If a tag is deployed in such an environment, the reflections from the tag are cluttered with the reflections from many other objects. To detect the tag in such a scenario, the back-scattered signal from the tag has to be isolated from all the other reflections. One way to achieve this is by doing cross-polarization~\cite{nolan2021ros, trzebiatowski202060} where the polarization of the tag reflected signals is different from the rest of the reflections. However, using this approach requires changes in the radar hardware making it non-compliant with COTS hardware. Inspired by existing work in RFID, we tackle this problem by modulating the back-scattered signal from the tag with a characteristic signature of the tag. 



\subsection{What kind of modulation to use on the tag?}\label{sec:mod_techniques}
A basic technique to modulate the tag's backscatter signal is to periodically turn the tag ON and OFF and create a characteristic pattern. In the ON state, the tag retro-reflects the incident signal and in the OFF state, the tag absorbs the incident signal. The choice of the time period of the modulation in comparison with the chirp duration has different design tradeoffs. To understand it, we divide the modulation methods into 2 frequency regimes a) when the modulation frequency is small i.e. the tag state changes from ON to OFF state over a span of multiple chirps(figure \ref{fig:in_vs_inter_chirp}a), b) when the modulation frequency is large i.e the tag state changes from ON to OFF state multiple times within a single chirp duration (figure \ref{fig:in_vs_inter_chirp}b).


\textbf{a)Low Frequency or Doppler domain Modulation (Inter-chirp):}
For the inter-chirp modulation case, the tag is in a fixed state(either ON/OFF) for multiple consecutive chirps and so we have to look at multiple chirps to observe the tag modulation. Here, the modulation frequency acts as the characteristic signature to identify the tag. Since the ON/OFF modulation is happening over a span of multiple chirps, the phase of the reflections from the tag change once over multiple chirps. This phase change creates a characteristic sinc-shaped pattern in the doppler domain of the Range-Doppler FFT (figure~\ref{fig:in_vs_inter_chirp}a). This technique is implemented in~\cite{soltanaghaei2021millimetro}, where they generate multiple sinc templates corresponding to different frequencies. During runtime, these templates are matched in the range-doppler domain, by searching over all range bins, to determine the frequency used in the modulation as well as the distance of the tag. Multiple non-overlapping frequencies are used to scale it to multiple tags. 

\textbf{b)High Frequency or Range Domain Modulation(In-chirp):}
In contrast to Inter-chirp modulation, using a high-frequency switching allows multiple cycles of the square wave to complete within a single chirp duration(figure \ref{fig:in_vs_inter_chirp}b). In this case, the tag's backscatter signal is shifted according to the frequency of the square wave in the range domain. For instance, if a tag's backscatter signal appears at the frequency $\Delta $ in range FFT, the high frequency $f_m$ modulation will create harmonics at $\Delta + n*f_m$ where $n$ is an odd integer. The presence of such frequency patterns acts as an indication that the received signal contains a tag backscatter signal and a template matching technique can help identify the tag's presence. This technique can be scaled to multiple tags by assigning different modulation frequency values to different tags. Note, that a very promising feature of In-chirp modulation over its Inter-chirp counterpart is low latency, as it allows operation within a single chirp.

\subsubsection{Practical implications of Inter-Chirp vs In-chirp modulation:} Although square wave modulation techniques presented above help us in uniquely identifying the tags, there are some practical challenges associated with them. The problem with the plain square wave modulation techniques is their unreliability in the presence of clutter due to environmental reflections. For instance, In the case of Inter-chirp modulation, a strong reflection from a moving object may fall into the doppler bin of a sinc template corresponding to some tag A and can be mis-detected as some tag A that is not even present in the environment(False positive). Note that the strong reflection from the vehicle is practical given the large size of the vehicle compared to the tags. 

On the other hand, the In-chirp modulation method enables the lowest possible latency by completing the modulation within a single chirp. This way we achieve our original requirement of low latency for COTS compliance and operational flexibility. However, the In-chirp modulation scheme is still plagued by reliability issues, similar to Inter-chirp modulation. The template formed in the range-domain can be confused by the overlap of the harmonic peaks of the square wave from two different tags located at different distances from the radar. Although In-chirp modulation decreases the latency issue, a collision between two tag reflections remains an issue. To tackle this challenge, \name designs a novel spread spectrum modulation scheme that maintains the low latency of in-chirp modulation and guarantees a reliable operation. 

\textbf{Effect on the reflected signal power:} \textcolor{\hlcolor}{
While performing the periodic ON-OFF modulation at the tag, the radar reflections from the tag contain a periodic pattern of signal and gaps. The gaps represent no energy is being reflected from the tag during the gap duration. Over a full ON-OFF cycle, half of the time there is no energy reflected from the tag. So, there is a 50\% loss in energy (3 dB reduction) in the reflected power because of the ON-OFF modulation. In the case of Inter-chirp modulation, this periodic pattern appears over a duration of a frame with half the number of chirps having no energy reflected from the tag. For In-chirp modulation, the ON-OFF pattern can be observed in a single chirp with half of the duration of each chirp having no reflected energy from the tag. In both the cases, there is a 3 dB reduction in the energy received from the tag and results in exactly the same SNR degradation for both In-chirp and Inter-chirp cases.}

\textbf{Effect on Tag ranging accuracy and resolution:} \textcolor{\hlcolor}{As seen previously, the ON-OFF modulation creates harmonics in the tag reflected signal either in the range-domain or in the doppler domain depending on the time scale of the ON-OFF switching frequency. These harmonics fall into an FFT bin exactly separated by the switching frequency away from the range bin of the tag. In order to range the tags accurately, we need to find the bin in which the harmonic is falling and subtract the switching frequency. So, the accuracy of the tag ranging depends on how accurately we can estimate the harmonic frequency. Estimating the harmonic's frequency depends only on the number of fft points and is not dependent on the ON-OFF modulation frequency. This implies the accuracy of localizing the tag is the same for both In-chirp and Inter-chirp modulation. At the same time, the resolution achieved in separating two tags depends only on the radar bandwidth and doesn't depend on Inter-chirp vs In-chirp modulation.}

\begin{figure}[t!]
    \centering
    \includegraphics[width=0.7\linewidth]{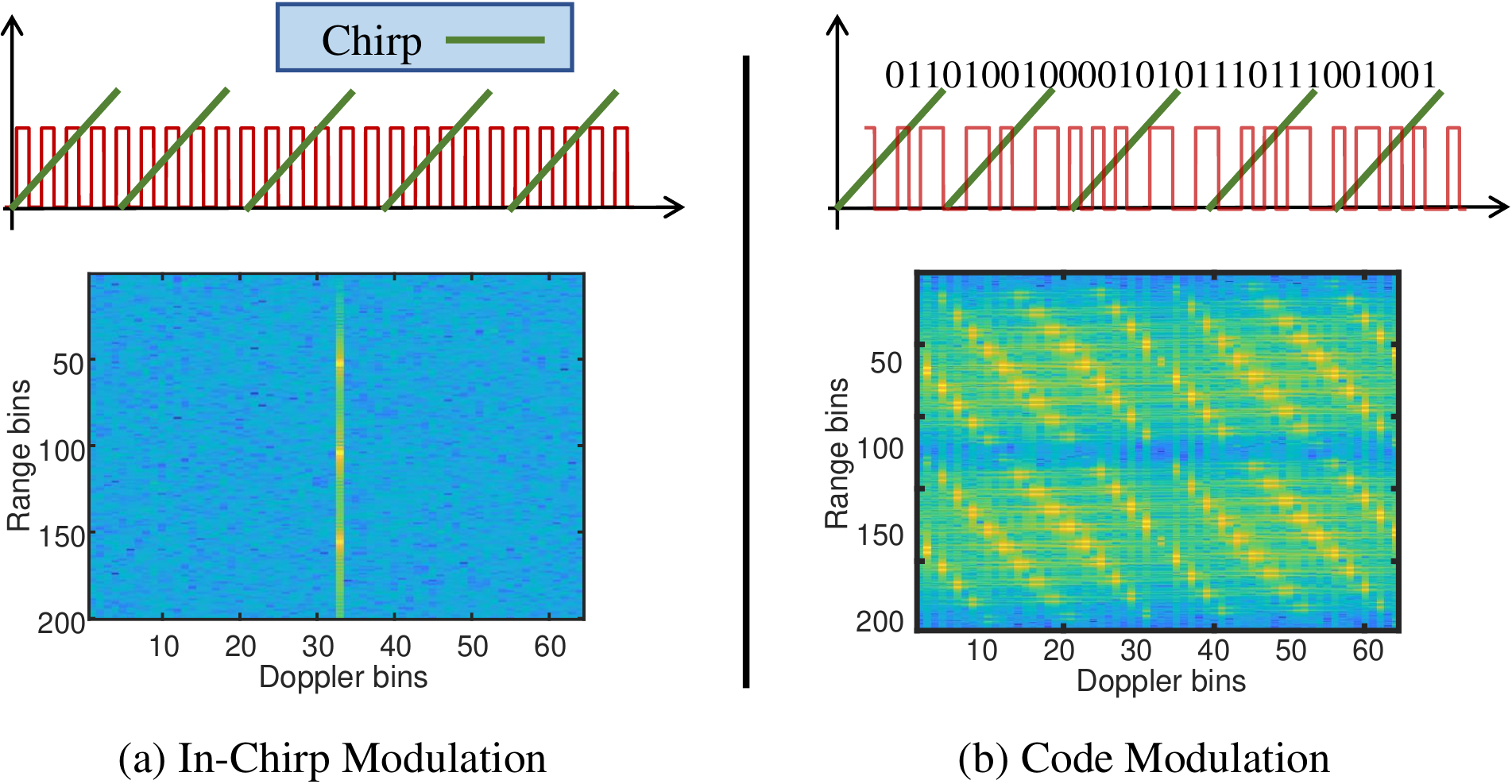}
    \caption{{Figure showing a) Low latency In-chirp square wave modulation and its Range-Doppler FFT showing the harmonics occupying only a few range bins b) Low latency Spread spectrum code modulated on every chirp to resolve collisions between tags. Range-Doppler FFT shows the code encodes tag modulation across the whole range bins providing reliable detection of tags.}}
    \label{fig:code_modulation}
\end{figure}

\subsection{\name's Modulation: Spreading Spectrum Code }\label{subsec:CDMA_code}
\label{subsec:distinguish tags}
Till now, we have seen that the plain square wave Inter-chirp and In-chirp methods are unreliable in Doppler and Range domains due to collisions between multiple tag reflections and reflections from the other objects in the environment. Here we propose an In-chirp method that is resilient to these collisions in the Range domain and can support the simultaneous operation of multiple tags. Our solution is inspired by the CDMA(Code Division Multiple Access) style of communication wherein multiple wireless users in the network make use of spreading codes to operate simultaneously. The spreading codes of two users are nearly orthogonal to each other and the user's signals can be separated from each other. Following this ideology, we design spreading codes that can be associated with our tags, allowing simultaneous multi-tag operation. 

Each of the spreading codes is an N-bit sequence that takes values in  {+1,0} with good auto-correlation and very low cross-correlation with other spreading codes. These spreading codes continuously repeat on the tag and modulate the backscattered signal. In this way, a spreading code serves as the characteristic property of the tag and helps in isolating a tag's backscatter signal from the reflections due to other tags as well as other objects in the surroundings. To see how the codes help in distinguishing two different tags, let us consider that two tags are using codes $C_1[n]$ and $C_2[n]$. Reflected chirps from the tags contain code1 $C_1[n]$and code2 $C_2[n]$ periodically repeating on them giving rise to  $[C_1[n] C_1[n]....C_1[n]] + [C_2[n] C_2[n] ....C_2[n]]$ at the receiver. Upon cross-correlating the received signal with $C_1[n]$, the resulting correlation is dominated by code1 because $C_2[n]$ is very weakly correlated with $C_1[n]$. Also, the cross-correlation looks like a series of impulses with $k$ peaks where k is the number of code repetitions in a chirp duration. Similar observations can be made when correlating with the code $C_2$. The periodicity of these peaks in the cross-correlation helps us identify the presence of a particular code. 
\begin{align*}
    \text{Cross-correlation} &= C_1[n]*[C_1[n] C_1[n]....C_1[n]] + C_1[n]*[C_2[n] C_2[n] ....C_2[n]]\\
    &\approx \sum_{l=0}^{l=k} \delta[n-lN] + 0
\end{align*}

To incorporate these spreading codes onto the tag's backscattered signal, the modulation signal $m(t)$ has to be modified to contain the N-bit spreading code. Each bit modulates the phase of the square wave pattern. A bit of value 1 corresponds to the ON to OFF transition of the tag with period $\frac{1}{f_m}$. Similarly, a bit of value 0 corresponds to OFF to ON transition of the tag with period $\frac{1}{f_m}$. The modulation signal is generated by appending the phase-modulated square waves corresponding to each bit as shown in figure\ref{fig:code_modulation}b. Each bit has a period equals to $T_{bit} = 1/f_m$ and occupies a bandwidth of $\frac{2}{T_{bit}} = 2f_m$.  The bit sequence $c(t)$ occupies is made up of N bits and occupies a total bandwidth of $2f_m$. Then the modulation signal $m(t)$ is the bit sequence $c(t)$ multiplied by the square wave of frequency $f_m$. We observe that $m(t)$ contains copies of $c(t)$ centered at $f_m$ and $-f_m$ thus occupying a total bandwidth of $4f_m$.
\begin{align*}
    m(t) &= c(t)\times \text{square wave of freq } f_m \approx c(t) \times \frac{2}{\pi}cos(2\pi f_m t) \\
    &= \frac{c(t)}{\pi}\times [ e^{j2\pi f_m t} + e^{-j2\pi f_m t}] 
\end{align*}

\textbf{Choice of Modulation frequency:} So far we have seen that the modulation signal depends on the frequency $f_m$ and the code $c(t)$. Here we explain how to choose the modulation frequency $f_m$ and its consequences on the backscattered signal spectrum. As explained in the previous section, the modulation signal $m(t)$ has 2 copies of $c(t)$ centered at $f_m,-f_m$. Each copy has a bandwidth $f_m$ and $m(t)$ occupies a total bandwidth of $4f_m$. The modulation constraints arise because of the following conditions: (a) Each backscattered chirp contains $k$ repetitions of the code and (b) the radar's Analog to Digital converter (ADC) has a finite sampling rate of $f_s$. 
To meet the first condition, the bit duration in each code has to be adjusted to ensure $k$ code repetitions in a chirp duration $T_c$. Recall that a code contains N-bits each of duration $\frac{1}{f_m}$, so each code occupies a length of $N/f_m$. Hence the chirp duration $T_c$ must be greater than $k$ code durations leading to the timing constraint given by the inequality \ref{eq:timing constraint}. 
\begin{equation}
    T_c > k\frac{N}{f_m} \implies f_m > k\frac{N}{T_c}\label{eq:timing constraint}
\end{equation}
Also for the radar to be able to capture the modulated signal, we have to ensure the spectrum of the modulating signal lies within the bandwidth of the ADC. Since the modulating signal $m(t)$ occupies $4f_m$ bandwidth, it leads to bandwidth constraint as given by inequality \ref{eq:bandwidth constraint}.
\begin{equation}
    4f_m < f_s \implies f_m < \frac{f_s}{4}\label{eq:bandwidth constraint}
\end{equation}
So, for a given chirp duration and the radar's sampling frequency, the modulation frequency $f_m$ has to be chosen to satisfy both the timing and bandwidth constraints.

\textbf{Choice of Spread Spectrum Codes:} Here we explain how to choose the spread-spectrum codes for each of the tags. To scale \name to sensing applications involving multiple tags, we assign a unique spread spectrum code(a pseudo-random bit sequence) to each \name tag. To have good autocorrelation and cross-correlation properties between the codes, we use Gold codes\cite{gold1967optimal,gold1968maximal} that are widely used in the GPS\cite{spilker1978gps} and the CDMA\cite{dinan1998spreading} standards. 

There exist Gold code sequences with different lengths and a natural question to ask here is to ask how to choose the length of Gold sequences and how many \name tags can be simultaneously used for the choice of Gold codes? This is akin to asking what is the capacity of the near-orthogonal codes used in a spread spectrum communication system. To answer this, we briefly explain the way Gold codes are generated and their properties. Gold codes are generated by XORing the outputs of two m-bit Linear feedback shift registers. For an m-bit shift register, the length of the Gold code generated is $2^m -1$. In total, $2^m+1$ unique Gold code sequences are generated by loading the shift registers with different initial conditions. Since these codes are not perfectly orthogonal, not all the unique code sequences can be used. The ratio between auto-correlation of a gold sequence and cross-correlation with other gold sequences is a measure of orthogonality for these pseudo-random sequences. So, when multiple tags are in use, the cross-correlation between the codes starts to approach the same level as the auto-correlation and the orthogonality measure between the codes starts diminishing. As long as the cross-correlation\cite{kim1988system} of a code with other codes/tags being used is below the auto-correlation, we can continue adding more tags to the system. We notice that the Gold codes have a special property that the cross-correlation between two sequences takes just 3 different values and the maximum of the 3 values is $2^\frac{m+1}{2} + 1$. The peak value of the auto-correlation for a gold-sequence is $2^m-1$ and for every new tag/code introduced into the system, the cross-correlation starts increasing almost by a factor $2^\frac{m+1}{2} + 1$. So, the total number of simultaneously usable codes($N_c$) to maintain auto-correlation peak more than the cross-correlation is given by  $\frac{2^m-1}{2^\frac{m+1}{2}+1}\approx 2^\frac{m-1}{2}$. For instance, if we choose m = 3 bit shift register, 2 tags can be simultaneously supported. It can be extended to support a higher number of tags by selecting a higher length shift register.

\textbf{Compatibility with radar modes:} A unique advantage of using the In-chirp spread spectrum modulation is its compatibility with all modes of radar. The single chirp operation not only provides low latency but also allows a seamless operation with analog beamforming radars. In analog beamforming radars, a chirp is directed to different angles to scan the entire scene. If the modulation requires multiple chirps to operate, it would be completely missed by an analog beamforming radar. Hence, the unique design of \name's modulation fulfills the requirement of compatibility with all radar modes.



\begin{figure*}[t!]
    \centering
    \includegraphics[width=0.9\linewidth]{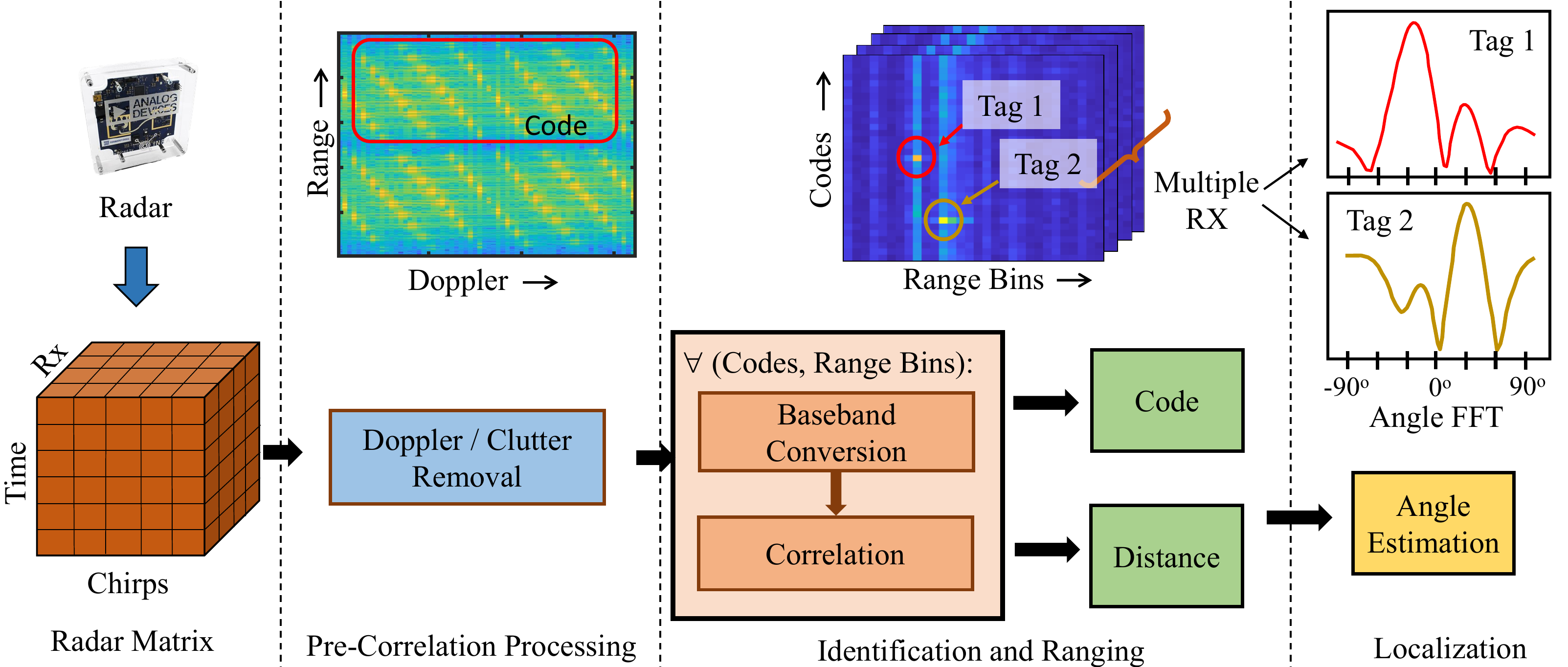}
    \caption{{Radar Processing Overview: The received signal from the radar is preprocessed to remove doppler and static clutter. The next step jointly solves for identifying the correct code and distance from the radar. The correlation values for the identified tags across multiple receivers are used to estimate the angle of each tag, thus providing the locations.}}
    \label{fig:Algorithm overview}
    \vspace{-10pt}
\end{figure*}

\section{Detection and Localization of \name}
In the previous sections, we discussed the two primary characteristics of \name -- Retroreflective design and code-based modulation. These characteristics create a tag design that can be detected and uniquely identified by any automotive radar. In this section, we build algorithms(overview in Fig. \ref{fig:Algorithm overview}) that can detect one or more tags and then localize them accurately i.e. find the distance and angle of the tag with respect to the radar. 

\subsection{Identifying and ranging the tag using cross-correlation}
\label{sec:section4.1}
Recall that the chirps reflected from the tag contain the modulating signal $m(t)$ in the radar's received signal. Upon dechirping the received signal at the radar, the dechirped signal contains the tag reflection $D_{mod}(t) = m(t)cos[2\pi\Delta t]$ where $\Delta$ is the frequency corresponding to the distance of the tag from the radar. We decompose the tag reflection as follows:
\begin{equation}
\begin{split}
   D_{mod}(t) &= c(t) \left(\frac{2}{\pi} cos[2\pi f_m t] cos[2\pi\Delta t]\right)  \label{eq:tag backscatter} \\
              &=\frac{c(t)}{2\pi}(e^{j2\pi (f_m+\Delta)t} + e^{-j2\pi (f_m+\Delta)t}\\
              &+ e^{j2\pi(f_m-\Delta)t} + e^{-j2\pi(f_m-\Delta)t} ) \\
\end{split}
\end{equation}
As shown in equation \ref{eq:tag backscatter}, the tag's reflected signal at radar contains multiple copies of code $c(t)$ centered at different frequencies $f_m+\Delta,f_m-\Delta,-f_m+\Delta,-f_m-\Delta$. Given the received samples at the radar $D_{mod}(t)$, how do we identify that the spectrum contains the code $c(t)$? The key challenge is that we have two unknowns for each tag: the distance of the tag (captured in $\Delta$) and the identity of the tag (capture code in $c(t)$). Furthermore, note that the received signal also consists of multiple reflections from the environment. 

We analyze the spectrum of the received signal to jointly solve for the tag-radar separation and the identity of the tag. We leverage the insight from before, that code modulated by the tags are orthogonal to each other (section \ref{subsec:distinguish tags}). So a cross-correlation with the code itself would reveal if the code exists or not. Specifically, if we compute the cross-correlations of the received samples with the correct code then we should observe high cross-correlation with periodically repeating peaks in it. But a straight-forward cross-correlation of $D_{mod}(t)$ with the correct code $c(t)$ do not result in a very good correlation due to the following reason: there are 4 different copies of the code in $D_{mod}(t)$ with a frequency offset term on each of them that corrupts the correlation.

The primary requirement to obtain a good cross correlation between the received signals and code is to eliminate the frequency offsets. Note that frequency offsets are not known as we don't know the $\Delta$, which is proportional to tag-radar distance. To overcome this challenge, we jointly solve for both $\Delta$ and the code that is present on the backscatter signal by iterating over the possible range of values that $\Delta$ can take.

\begin{figure*}[t!]
    \centering    
     \includegraphics[width=\linewidth]{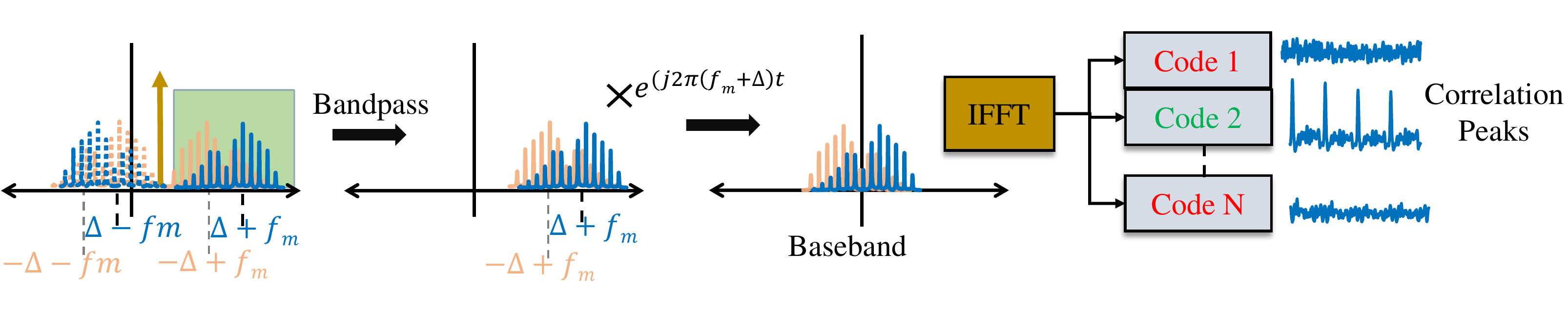}
    \caption{Correlation: The received backscattered signal from the tag contains 4 copies of the code in the frequency domain due to aliasing. The component at $\Delta + f_m$ is retrieved using bandpass filtering and then converted to baseband. The time-domain version of the signal is then correlated with all the codes in the codebook for identifying the tag. The right plot shows the gram matrix for correlation between different codes for a 33 gold code sequence.}
    \label{fig:cross_correlation}
\end{figure*}

To remove the effect of the frequency offset,  we process the samples of $D_{mod}(t)$ before taking cross-correlation in the following way: The first step of this processing is to filter out the terms in the equation \ref{eq:tag backscatter}that would lead to bad correlation. Note that the code copy centered at $f_m+\Delta$ has spectrum in the frequency range $\Delta$ to $2f_m+\Delta$ whereas the code copy centered at $f_m-\Delta$ has spectrum in the frequency range $-\Delta$ to $-2f_m-\Delta$ as shown in Figure~\ref{fig:cross_correlation}. These two copies have no common frequency content and are perfectly separable in frequency. Also, we observe that the code copies centered at $f_m+\Delta, f_m-\Delta$ are in close proximity with each other. Similarly, code copies centered at $-f_m+\Delta,-f_m-\Delta$ are close to each other. So, first we pass the samples of $D_{mod}(t)$ through a bandpass filter of bandwidth $2f_m$ as shown in figure~\ref{fig:cross_correlation} to eliminate the code copies centered at $-f_m+\Delta, -f_m-\Delta$ and retain the copies centered at $f_m+\Delta, f_m-\Delta$ in the filtered signal. 
\begin{align}
    \text{Filtered Signal} = c(t)[e^{j2\pi (f_m+\Delta)t} + e^{j2\pi (f_m-\Delta)t}]\label{eq:Right half filtered signal}
\end{align}
In the next step, we remove the frequency offset factors in the filtered signal by shifting the code copy centered at $f_m+\Delta$ to zero frequency. This is achieved by multiplying the filtered signal with $e^{-j2\pi(f_m+\Delta)t}$ leading to $x(t)$:
\begin{align}
    x(t) = c(t)  + c(t)e^{j2 \pi(-2\Delta)t}. \label{eq:Right half spectrum}
\end{align}

\begin{figure}[t!]
    \centering
    \includegraphics[width=\linewidth]{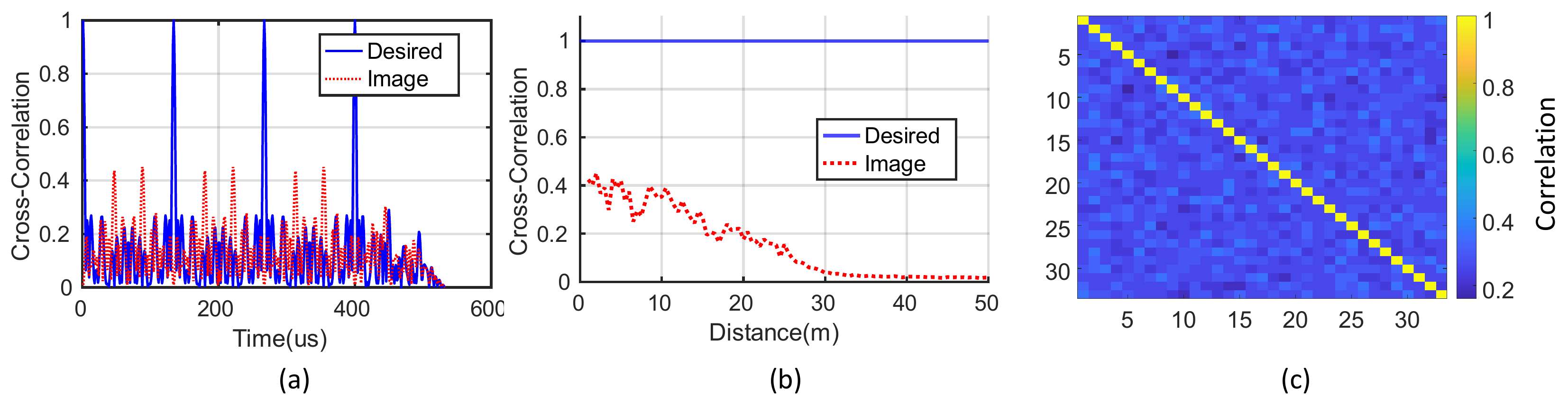}
    \caption{{(a) Normalized cross-correlation for 2m tag to radar separation. (b) Peak value of cross-correlations as a function of distance. Image signal's contribution to the cross correlation reduces gradually as tag to radar separation increases (c) Gram matrix showing the auto-correlation and cross-correlation of 31 bit Gold codes. }}\label{fig:cross-corr}
\end{figure}


The resulting signal $x(t)$ contains the desired signal $c(t)$ and an additional term with a frequency offset of $2\Delta$ which we call as the image signal. When $x(t)$ is correlated with the correct code sequence $c(t)$, the first term $c(t)$ results in a periodic impulse train. The image signal $c(t)e^{j2 \pi(-2\Delta)t}$ contains a frequency offset of $2\Delta$ that increases with the tag to radar separation. Because of the frequency offset, its cross-correlation becomes smaller as the tag to radar separation increases. In figure \ref{fig:cross-corr}, we empirically show that the cross-correlation contribution from the image signal becomes smaller as $\Delta$ increases. 
Similarly, when $x(t)$ is correlated with an incorrect code, it leads to a poor cross-correlation. In this way, we can cross-correlate $x(t)$ with all the possible codes and find the set of codes that result in an impulse train like cross-correlation and declare the codes that are present in the backscatter signal. We store the highest peak of the cross-correlation in a 2D matrix over all the range bins and the codes. The highest value in this 2D matrix jointly gives us the tag to radar separation and the code that is present in the backscatter signal.

\textbf{Performance in doppler} Here we present the impact of the doppler effect caused by the moving vehicle on the joint decoding. The fast movement of a vehicle during a chirp duration causes the received chirp to be slightly shifted in a frequency proportional to the carrier frequency and the velocity of the vehicle (doppler effect). This shift in frequency manifests as a shift in the code spectrum by few bins in the range FFT. Since our joint decoding algorithm iterates over all possible shifts, we will still be able to identify the tag. But the shift in the spectrum introduces the error in the estimated tag to radar separation. Also, as our decoding scheme does not require template matching in the range doppler domain, the radar never confuses between doppler arising from a vehicle and the code spectrum. We simulate the doppler scenario for upto 100mph speeds(see evaluation figure~\ref{fig:mobility_sim}) and show that doppler results in a maximum error of 2 metres.

\subsection{Combining across multiple chirps}
So far, we have shown how to detect tags by looking at the periodic peaks in the cross-correlation of the backscatter signal with a single chirp. The single chirp-based detection is important to detect the tag with beam-forming radars, which scan one direction at a time. However, there are many radars that use digital beamforming including the one used in our evaluation. In this section, we would present how we can combine multiple chirps to achieve much better accuracy.

Here we describe how to make our algorithm robust to the noise present in the backscattered signal. The strength of the tag's backscatter signal inversely varies with the fourth power of the tag to radar separation($\propto \frac{1}{d^4}$). To put this signal power variation into perspective, if the tag to radar separation is doubled, the received signal power drops by 12 dB. This implies the reflected signal power can quickly hit the receiver noise floor and the characteristic peaks in the cross-correlations will be buried in the noise floor. In order to perform reliable long-range tag detection, we must enhance the  Signal to noise ratio(SNR) of the received signal. 

For this reason, our algorithm combines radar samples from multiple received chirps to make the periodic peaks in the cross-correlation stand out from the noise. Typically radars have a gap time between two consecutive chirps and so we fill the gaps between the chirps with zeros and append multiple chirp samples. Then we perform cross-correlation with the code sequence to detect the peaks and thus identify the code that is present. Cross-correlation on these appended samples is equivalent to averaging over noise samples to reduce the noise power which gives us the processing gain. So, combining $L$ chirps reduces the noise power by a factor of L and results in a processing gain of $10 \log_{10}(L)$dB. We set a threshold based on the peak-to-noise ratio for detecting the presence of a particular code. Any correlation value greater than the threshold is marked as a positive detection.


\subsection{Latency of operation} \textcolor{\hlcolor}{As discussed previously, processing samples from multiple chirps enables us to perform long-range tag detection. The processing gain or the improvement in SNR logarithmically increases with the number of chirps processed together. So, to detect the tags within a specific target distance from the radar, we will be able to decide how many chirps are to be processed coherently. On its own, the algorithm can work end-to-end with even a single chirp, but if more chirps are needed to be processed, a longer duration of radar samples can be accrued. Therefore, our algorithm offers the flexibility to choose the number of chirps for coherent processing. In section \ref{sec:evaluation}, we show results about the practical implications of longer processing times. Moreover, the latency of the system is independent of the number of tags present as the system always searches for the presence of all codes. The low latency operation of \name is specially useful in the case of radars that use beamforming-based scanning. Such radars send a single chirp in different directions in space to scan the environment. For these radars to detect the tag, it is extremely important to complete the modulation within a single chirp duration. }

\subsection{Localizing the tag}
\noindent As mentioned above, we jointly solve for the distance and identification of the tag by iterating over different codes and range bins. For localizing the tag, we also estimate the angle at which the tag is present. Distance and angle of tag provide us with the location of the tag. To estimate the angle of the tag, we leverage multiple receivers present on our radar. Based on the signal's angle of arrival (AoA), each subsequent receive antenna in the linear array receives an additional phase induced by an extra path length of $\frac{dsin(\theta)}{\lambda}$, where d is the separation between the two receive antennas and $\theta$ is the AOA. 

A signal reflected from any object appears as a constant sinusoidal tone in a particular frequency bin of the range FFT spectrum of the radar. To estimate the angle of arrival for this signal, the value at the corresponding bin is picked across multiple receiving antennas and an FFT over these values is used to determine the angle. However, in our case, the frequency spectrum due to the tag's back-scatter is spread over a bandwidth of $2f_m$ (section \ref{subsec:distinguish tags}). Thus, using a similar approach as described above is not possible in this case. 

To address this challenge, we use the signal obtained after the correlation with the correct code across multiple antennas. Note that the phase difference induced due to the difference in path lengths at different receivers is preserved during the cross-correlation. Taking angle FFT of correlations at multiple antennas allows us to pick the peak values in the angle FFT and in turn estimate AOA. The estimate of AOA together with the distance of the tag provides the location of the tag. An important thing to note is that the resolution of AOA is dependent on the number of available receivers on the radar (4 in our case). Current automotive radars contain large receive antenna arrays, which can be used to improve the angular resolution. 



\section{Implementation}
\label{sec:implementation}

\textbf{\name's tag hardware design:} The tag hardware is designed on a printed circuit board and can be split into two categories: (a) RF circuit design and (b) Control circuit.

\textbf{RF circuit:} The RF circuit of the tag is designed on a printed circuit board (PCB). The PCB consists of the Van Atta Array antennas that are connected by transmission lines and the RF switches to modulate the backscattered signals. The antennas used on the tag are patch antennas which are directional in nature. The antennas are designed to have a 50 ohm impedance over a frequency range of 24GHz to 24.3GHz to cover the 24GHz ISM frequency band. Each antenna has a 14 dBi peak gain and 100 degree horizontal field of view to support long range and wide field of view. The printed circuit board (PCB) is designed on a 20 mil Rogers 4003c dielectric as a substrate to minimize the RF trace losses. 

In order to modulate the information on the tag, we design the tag to switch between van-atta and absorbing states. To achieve two different states for the tag, we use an RF switch on each transmission line that connects a pair of antennas.  When the RF switch is turned on, the connection between the antenna pairs is established and the incident signal is retro-reflected, providing high RCS to radar. When the RF switch is turned off, the transmission line ends are \textcolor{\hlcolor}{terminated in a 50 ohm microwave resistor\cite{microwave_resistor}} and the incident signal is completely absorbed by the tag, providing lower RCS. In this manner, the RCS value could be controlled by using the switch, based on the information we send and the radar could discern the information from these fluctuations. In our implementation, we used the MASW-011105 RF switch from MACOM due to its low insertion loss(1.6 dB) at 24GHz and low DC power consumption of 5 uW~\cite{switch_datasheet}.

\noindent \textbf{Control ciruit:}
To control the switch on the tag we use an onboard PSoC 6 MCU (Microcontroller unit). We implement two Linear feedback shift registers (LFSR) on the MCU to generate the gold code sequence. The generated bits are outputted from a GPIO pin and are fed to the RF switch as the control signal. The LFSRs can be configured to use different length codes. In our implementation, we use 5-bit LFSRs to generate 32-bit codes.

\begin{figure}[t!]
    \centering    \includegraphics[width=\linewidth]{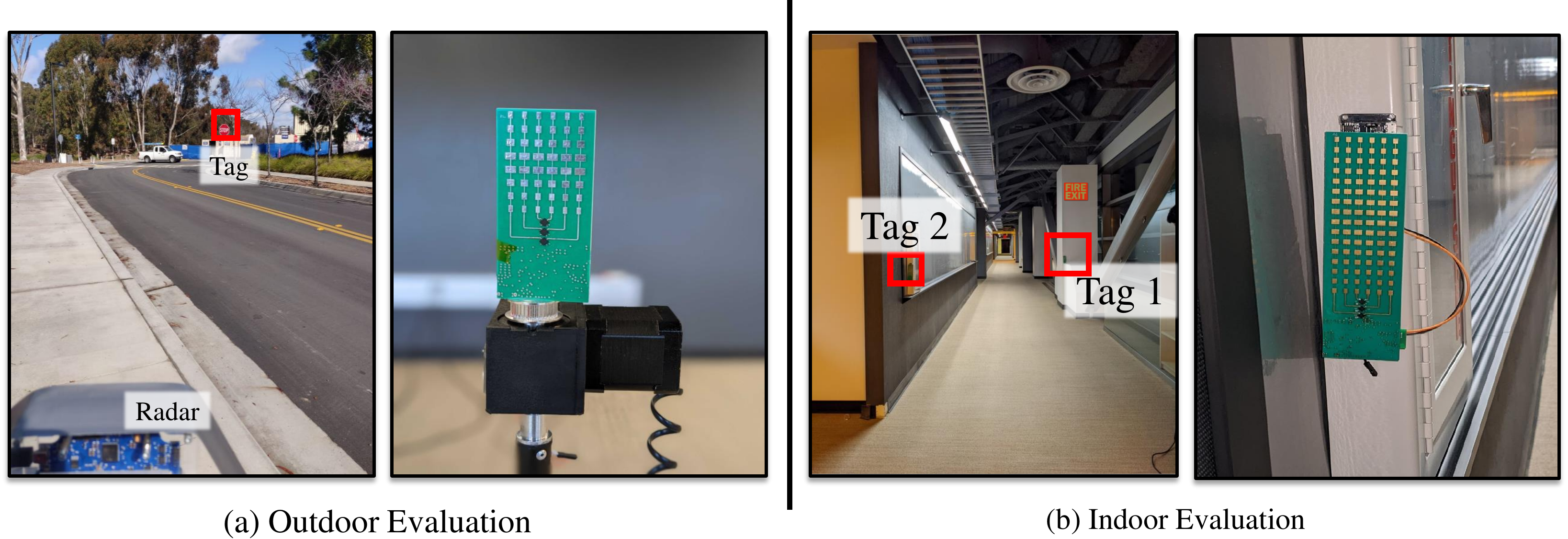}
    \caption{{Experimental setup for \name: (a) Shows an outdoor experimental scene for a tag placed on a stop sign. We use the 8-patch antenna design for outdoors. (b) Shows indoor experimental scene with multi-tag deployment in a multipath-rich environment. We use a 16-patch antenna design for indoor evaluation.}}
    \label{fig:exp_setup}
\end{figure}

 Figure~\ref{fig:exp_setup} shows the tags that we used to implement \name. We design 2 tags with 8 and 16 patches per antenna to use in outdoor and indoor applications.

\noindent \textbf{Tag Power consumption:}
The only active elements on our tag are the RF switches that modulate the backscattered signal. The power consumption of an RF switch consists of two components (a)Static and (b) Dynamic power. We will explain the power used by our tag in each of these components. Static power consumption of the circuit is caused by the leakage current in all the circuit components. In our implementation, the RF switches are operated at a supply voltage of 5v and draw a continuous current of $1\mu A$ from the supply thus consuming a static power of 5 $\mu W$. Dynamic power on the other hand is due to the energy dissipated in the charging and discharging of capacitors in a circuit. So, dynamic power depends on the rate of switching. In our implementation, we use commercial switches with a dynamic power of 140 $\mu$W at 250 kHz switching rate. Hence, in total, our tag consumes 145 $\mu$W of power but is mostly dominated by the dynamic power consumption. However, the dynamic power is also the function of the input power of the switch. Typically the commercial RF switches are designed to handle input power levels at 1 watt and thus they require transistors within them that can handle very high current. To handle high RF power, transistors with large gate sizes are used which significantly rises the input capacitance resulting in high dynamic power. In backscatter applications where \name is used, the incident RF signal power levels are of the order of a few $\mu W$, eliminating the need for large transistors. Hence by designing the RF switches on Application-Specific Integrated Circuits\cite{CMOS_mmwave}(ASIC) the dynamic power can be significantly reduced.

\noindent\textbf{Code Design and Radar parameters:}
To implement and evaluate our end-to-end detection system with \name, we use the 24 GHz DEMORAD radar platform from analog devices that provides chirp samples at a 1 MHz rate. It is a MIMO radar with 2 transmit and 4 receiver antennas. We use one transmit and 4 receive chains to collect the data and estimate the angle of arrival to find the tag's angular location. For the tag identification, we modulate the tag's backscatter signal using 31 bit Gold code sequences. These Gold code sequences are generated on the tag using a 5-bit linear feedback shift register. The code sequence is modulated at 250kHz modulation frequency $f_m$ to satisfy the bandwidth constraint as explained in section \ref{subsec:CDMA_code}. There exist 31 gold sequences of 31-bit length with good cross-correlation properties and our system is able to support 4 different tags that are located in the vicinity of each other. In our experiments, we configure the radar with an upchirp time of 496 $\mu$s and a gap time of 104 $\mu$s between two chirps. We fill this gap time with zeros in-order to get processing gain by combining samples from multiple chirps. We note that our implementation of \name is independent of these timing parameters and would work for any other choice of these parameters. Figure \ref{fig:exp_setup} shows our experimental setup.

\section{Evaluation}
\label{sec:evaluation}

\name design a smart mmwave sensing infrastructure, that enables applications in both indoor and outdoor settings. To evaluate \name we perform extensive experiments in both indoor and outdoor environments with realistic settings. Figure~\ref{fig:exp_setup} shows the two different settings where we evaluate \name. The evaluation showcases the suitability of using our system for smart mmwave sensing infrastructure. Our design of \name provides an operating range of more than 25m with a commercial radar, with 100\% reliable detection rate. Moreover, \name's retroreflectivity provides an FOV of more than 120 degrees which is vital for critical applications like autonomous driving and indoor object identification. With its unique spread spectrum code-based design, \name provides a reliable, scalable and low latency solution for augmenting mmwave sensing infrastructure.

In this section we would extensively evaluate how \name fulfills all the requirements mentioned in section~\ref{sec:intro}. We first evaluate the performance of a single \name tag for its applicability to the aforementioned indoor and outdoor applications. Next, we show the scalability of \name tags to multiple tags which is important for large-scale deployment. We also perform field tests of \name in the real-life driving scenarios and adverse conditions like fog and smoke. Finally, we provide some microbenchmarks to show the reliable performance of \name's spread spectrum-based coding design.

\begin{figure*}[t!]
    \centering
    \begin{minipage}{0.42\linewidth}
     \includegraphics[width=\linewidth]{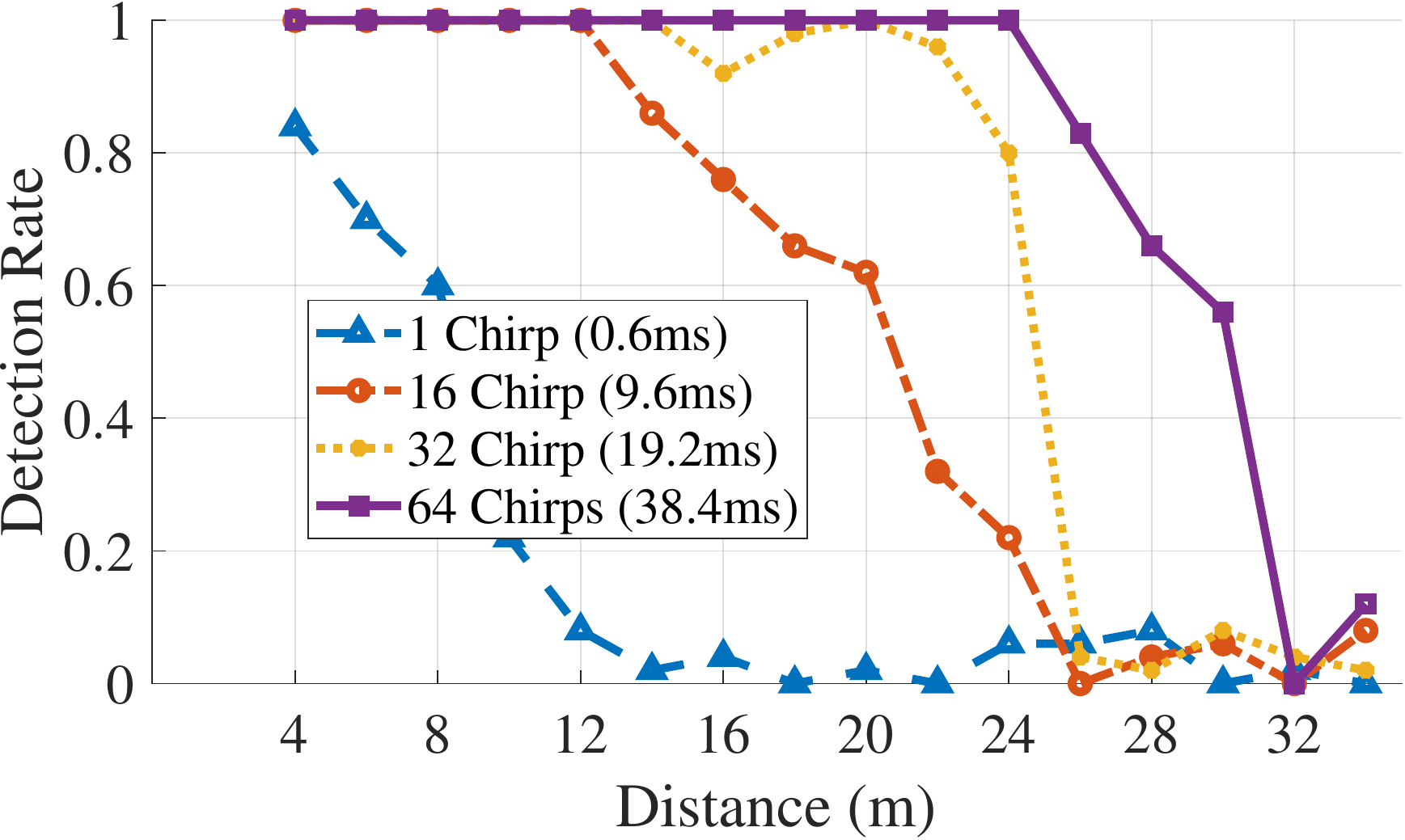}    
     \subcaption{Detection Rate with Distance}    
    \end{minipage}
    \begin{minipage}{0.42\linewidth}
     \includegraphics[width=\linewidth]{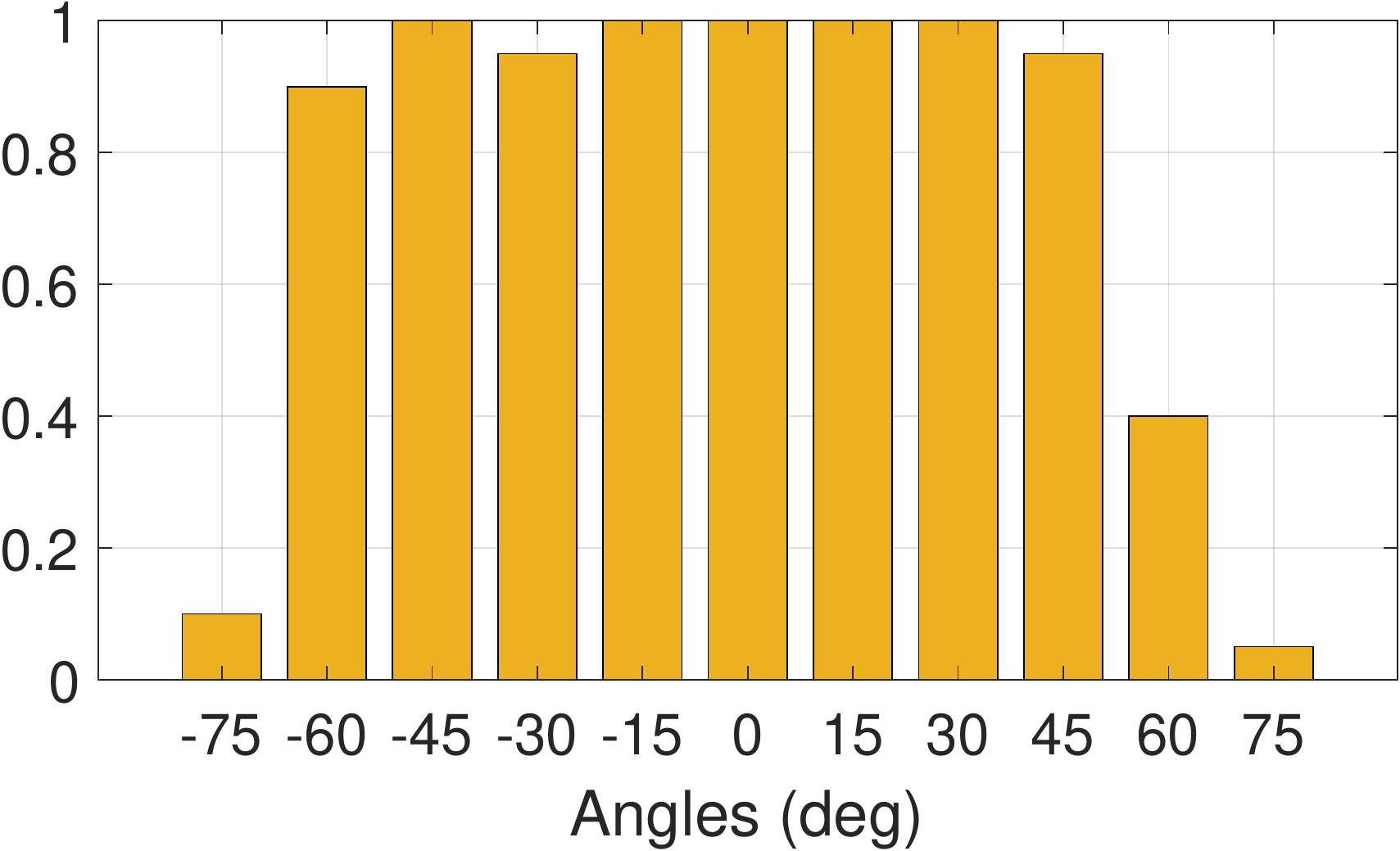}     
     \subcaption{Field of View}  
     \label{fig:tag_fov}
    \end{minipage}
    \vspace{3pt}
    \caption{{a) Detection Rate for different observation times. b) Detection rate for different angles of incidence showing wide field of view of our tags.}} 
    \label{fig:detection_range}
\end{figure*}

\subsection{\name's performance for a single tag}
\textbf{Field of View:}
One of the essential characteristics of \name is the wide Field of view. To understand the performance of \name for different angles of incidence we evaluate the detection rate of the tag for different angles. Figure \ref{fig:tag_fov} shows the detection rate of \name at different viewing angles at a distance of 5m from the tag. For each angle, we take 100 frame measurements to find the detection rate. It can be seen that \name can successfully retain its reliable detection performance for more than 60 degrees of incidence on each side, which makes it fit for use for even challenging cases for example reading a stop sign on a curved road or an exit sign from a long distance in a corridor.

\vspace{0.25cm}
\textbf{Range of operation:}
We have designed \name by considering the low latency requirements of the application. Due to its time-based correlation functionality, \name needs only a single chirp's data to perform all its operations. However, \name also provides flexibility to extend its operating range and improve the reliability, by coherently processing multiple chirps during runtime. Figure \ref{fig:detection_range} shows the detection performance of \name over a range of more than 30 m for different observation times. At each distance, 100 frame measurements are taken. Note, that the maximum range of detection heavily depends on the antenna gain of the tag~\cite{soltanaghaei2021millimetro} and the radar hardware. For instance, the maximum EIRP allowed at the 24GHz band is 20dBm, at which the detection range would further extend to around 45(130) m for 1(64) chirp combining. 
\textcolor{\hlcolor}{There are existing off-the-shelf automotive radars (Eg: K-MD2\cite{RF_beam}) that have a low receiver noise figure of 10dB(same as the radar we used in our experiments) while transmitting at a high transmit power of 20dBm. Since the noise figure of the radar remains the same while transmitting at higher power, it can enable a longer detection range when used with \name tags}. Our choice of 24 GHz radar was driven solely based on the non-availability of the 77GHz band components and our design has no dependency on the frequency of operation. 

\begin{figure*}[!t]
    \centering
    \begin{minipage}{0.32\linewidth}
     \includegraphics[width=\linewidth]{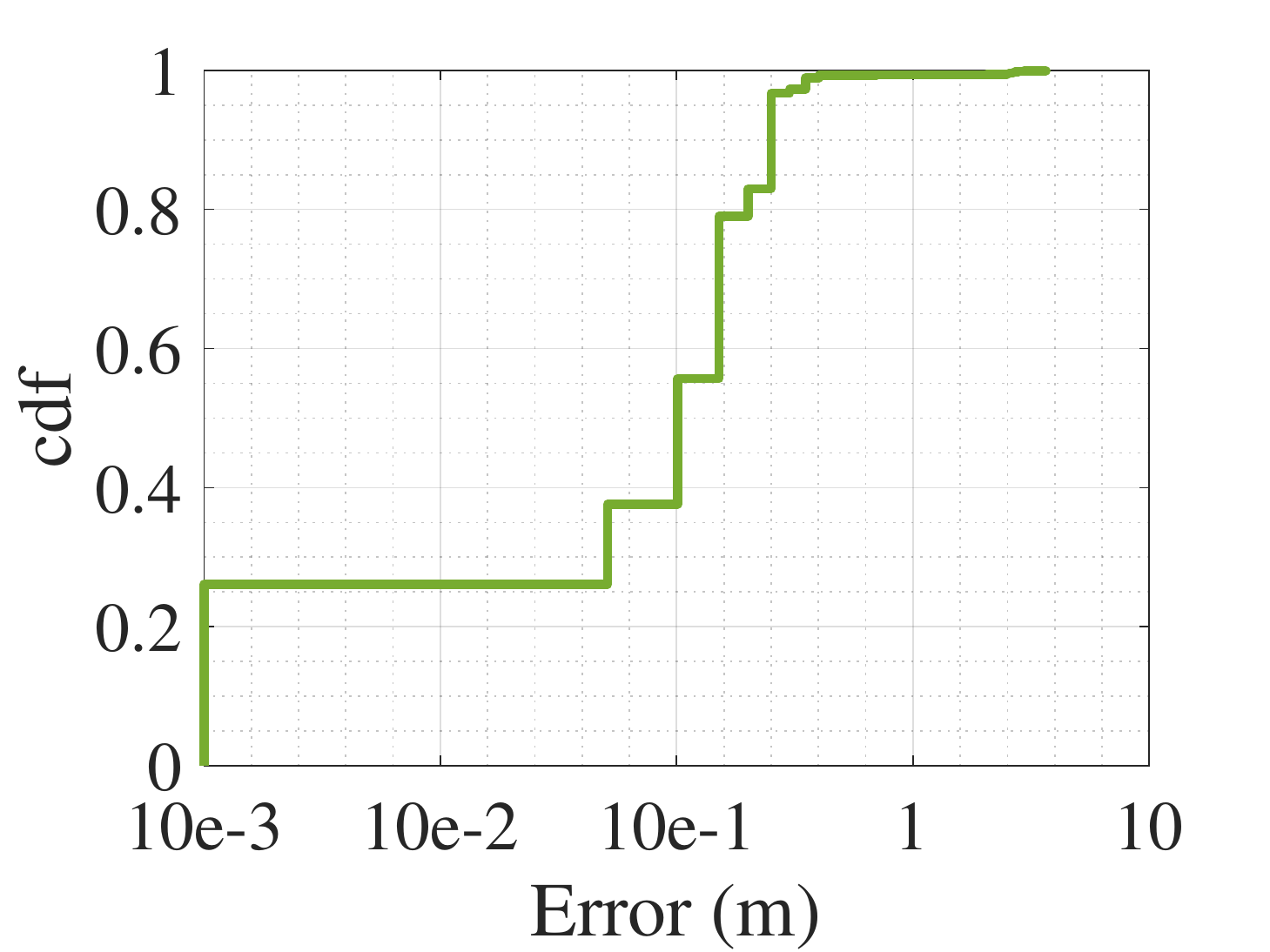}    
     \subcaption{Distance Estimation CDF}    
    \end{minipage}
    \begin{minipage}{0.32\linewidth}
     \includegraphics[width=\linewidth]{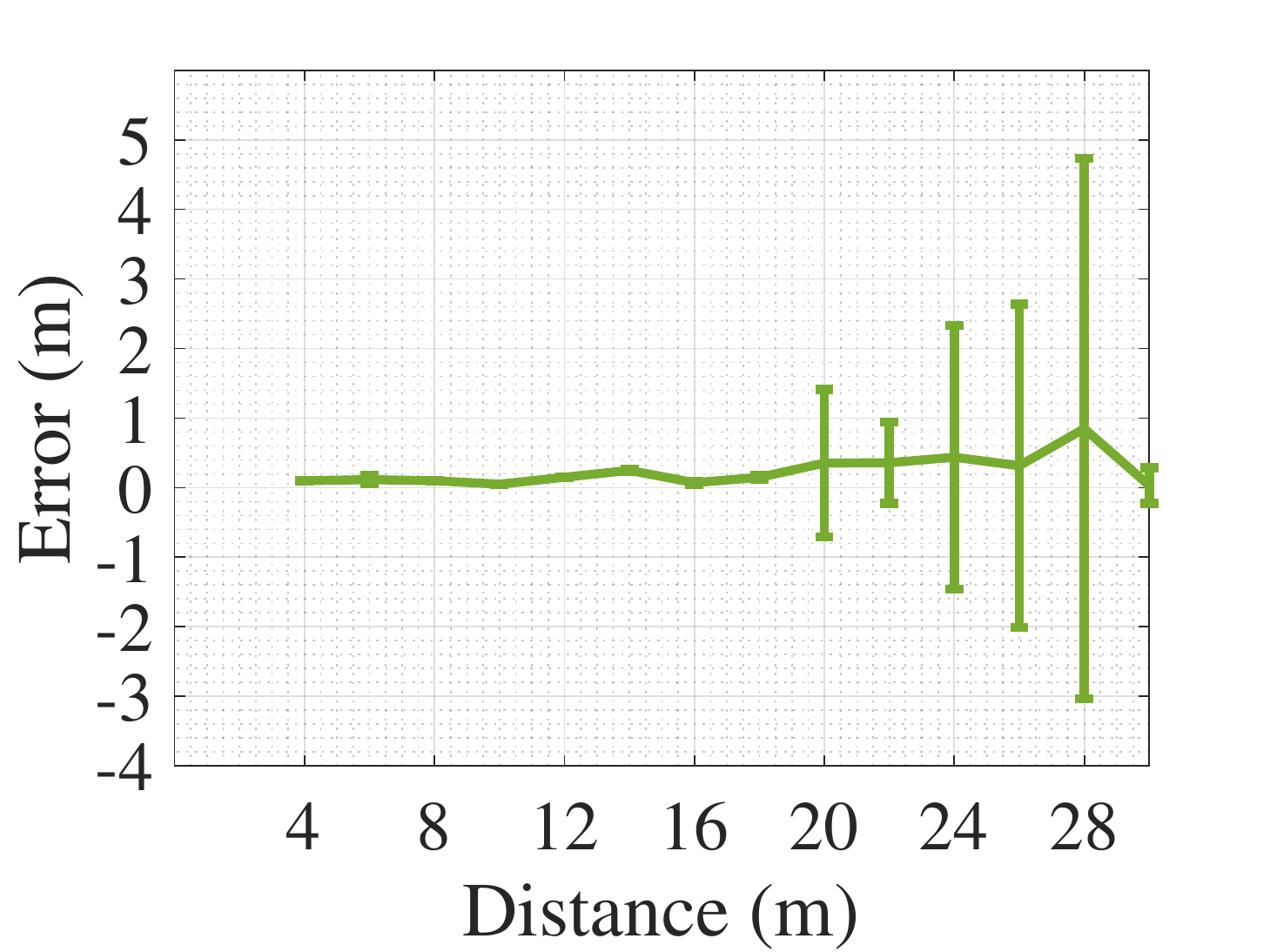}     
     \subcaption{Distance Estimation Error}  
     \label{fig:distance_wise_error}
    \end{minipage}
    \begin{minipage}{0.32\linewidth}
     \includegraphics[width=0.9\linewidth]{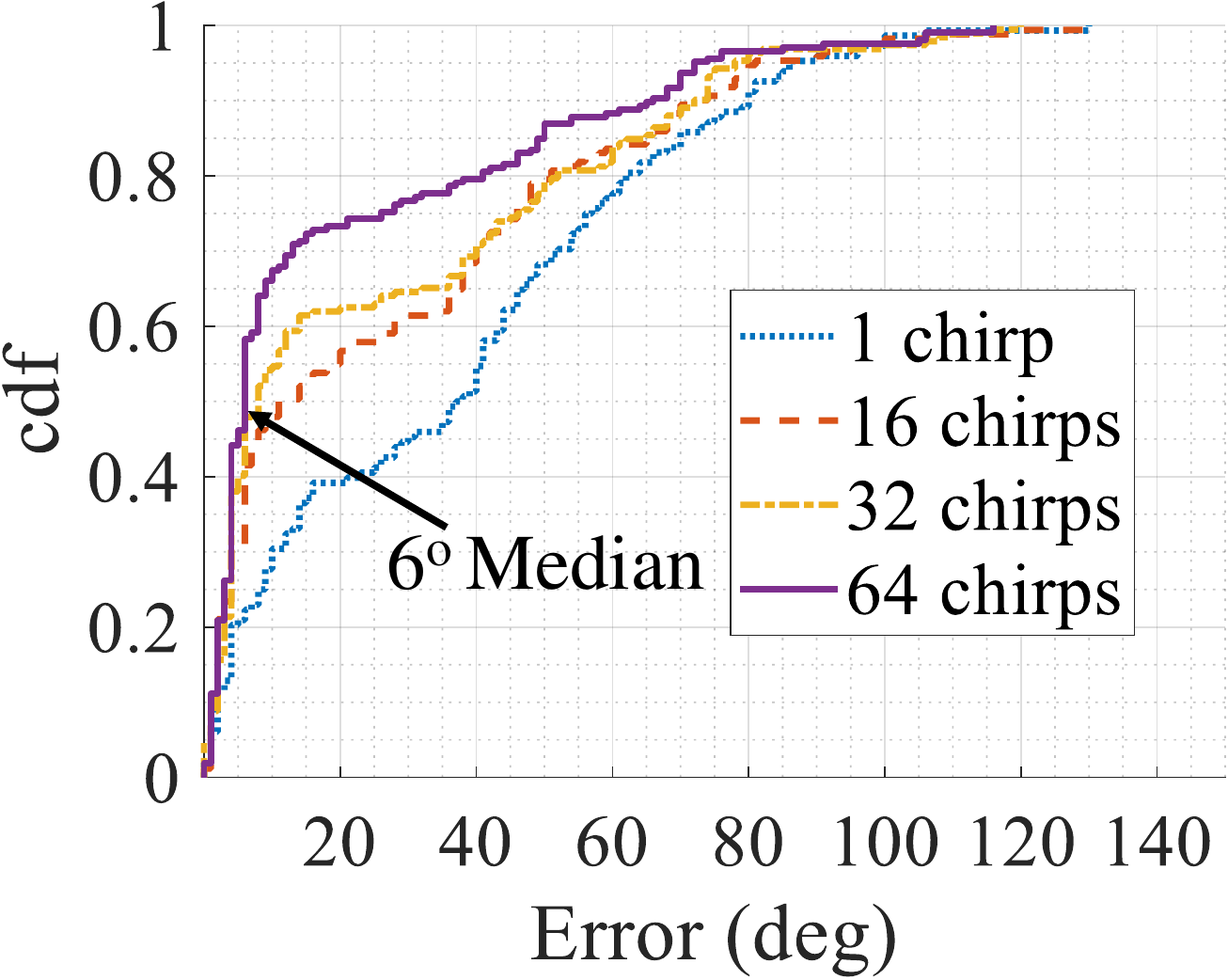}    
     \subcaption{Angle Error CDF}    
    \end{minipage}
    
    \vspace{3pt}
        \caption{(a-c) Localization performance of \name}

    \label{fig:localization}
\end{figure*}

\vspace{0.25cm}
\textbf{Localization:}
\textcolor{\hlcolor}{\name's algorithm provides the range and angle estimates of the tag, thereby localizing the tag in the environment. In this experiment, we place the tag at different distances and different angles from the radar in the outdoor environment. For each location, we take 100 measurements. Figure \ref{fig:localization} shows the error CDF plots of the tag's distance and angle estimation by our system. The median error of distance and angle estimation is 0.1m and 6 degrees, respectively. Figure \ref{fig:distance_wise_error} shows the error in distance estimation with increasing distance. Note that the error in measurement includes the error in ground truth measuring instrument, which is 1mm. The standard deviation of the localization error increases as we move farther from the radar because of the lesser SNR. We have used FFT-based distance and angle estimation in our system, but any super-resolution algorithm can also be used. However, for determining the distance of a particular tag, an FFT-based system would perform similar to any super-resolution algorithm. This is proven by the fact that past work~\cite{soltanaghaei2021millimetro} uses a super-resolution algorithm and achieves a comparable median error of 0.15 and 3 degree. The reason behind this is that the resolution of distance measurement does not affect the error of measurement (section 3). Hence, in our system, which has a distance resolution of 60cm, we can still achieve lower error simply by taking a larger point FFT~\cite{bhutani2019role}. In this evaluation, we take FFT points as 1984, which is four times the number of samples per chirp, i.e., 496. }


\begin{figure*}[t!]
     \includegraphics[width=\linewidth]{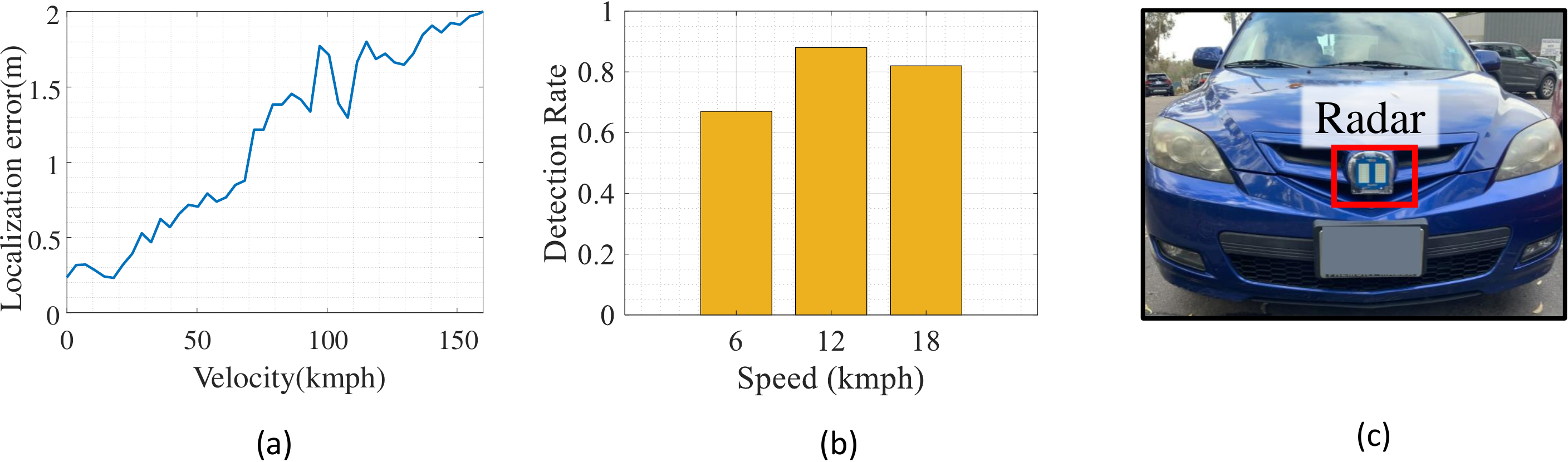}     
     
    \caption{(a) Simulation results of distance estimation error in case of high doppler. (b) The detection rate of \name in real-world experiments when the radar is mounted on car that runs at different speeds. (c) Experimental setup with radar deployed on the car. }
    \label{fig:mobility_sim}
    \end{figure*}

\subsection{\name's scalability to multiple tags}
\textcolor{\hlcolor}{
Scalability in terms of the number of tags that can operate simultaneously is a critical feature of \name. In this section, we evaluate \name's capability of supporting simultaneous tag operation. During the real-world operation, the correlation-based detector will find the correlation of all the codes in the codebook with the signal received. Unlike the single tag case where we could just look at the code corresponding to the maximum correlation value, we need to set a threshold of detection in multiple tag evaluation. The value of the threshold determines how many detections are made. If any correlation value exceeds the threshold, the corresponding code, and hence the tag, would be marked present. A lower threshold means more detections and possibly more false positives and vice versa for a higher threshold. Hence, to evaluate the performance of \name in the case of multiple tags, we use the AUC (area under the curve) metric, which is a standard metric used to measure a classifier's ability to distinguish between classes. The true positive rate is plotted against the false-positive rate for different threshold values. The area under the curve then determines the performance of the detector. }

\subsubsection{Detection performance and latency:} 
\begin{figure}[t!]
    \centering
    \includegraphics[width=0.8\linewidth]{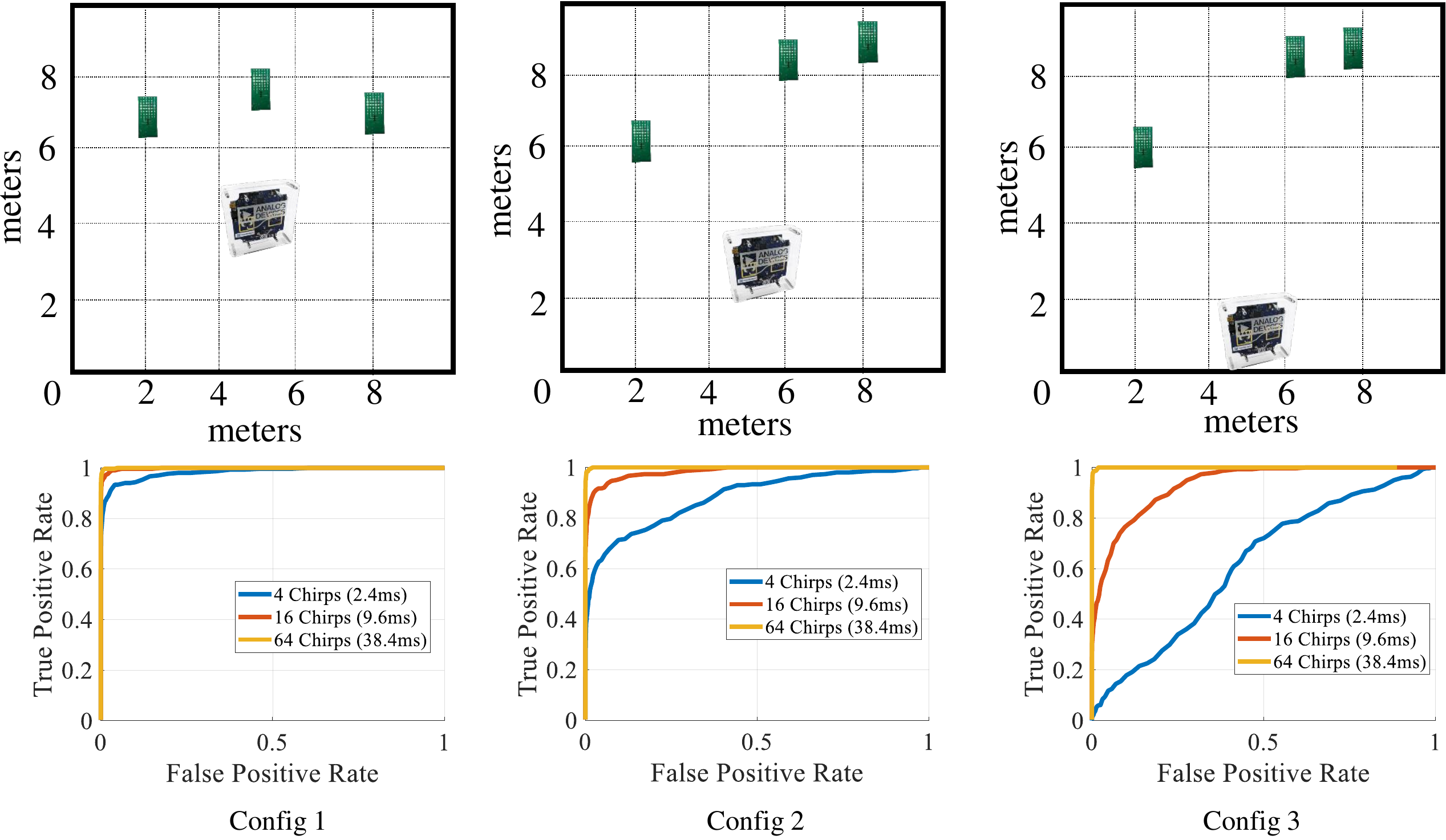}
    \caption{{ ROC curves for multi-tag experiments for different chirp combinations. More the curve towards the left the better }}
    \label{fig:multitag_3}
\end{figure}

\textcolor{\hlcolor}{For this experiment, we configure three tags with different codes and keep them at separate locations in three different configurations within a range of 10m from the radar. Different configurations cover cases of wide-angle separation (config 1) to closely spaced tags (config 3). The distance of the radar from tags also increases in each subsequent configuration. We capture 100 frames for the experiment in each configuration and plot the true positive rate against the false-positive rate. Figure \ref{fig:multitag_3} shows the curves for the performance of detection and identification in the presence of multiple tags. Table~\ref{tab:multi_tag_auc} shows the AUC calculated for each configuration for different chirp combinations. As expected, combining a large number of chirps (64) provides an almost perfect AUC. For smaller distances (config 1), using even a small number of chirps obtain a good AUC value. Even when the tags are closely placed, they can be clearly detected separately by \name. Note that the latency of operation is completely independent of the number of tags present in the environment. This is because we check for the presence of all the codes in all distance bins regardless of the number of tags. Every code and distance bin that gives a high correlation marks the presence of the tag (see figure~\ref{fig:Algorithm overview}). This experiment proves that \name maintains its high reliability even in presence of multiple tags. }

\subsubsection{Localization error:}
\textcolor{\hlcolor}{We conduct another experiment to analyze the effect of having multiple tags on localization accuracy. We place three tags in an indoor environment where the distance between tags varies from 0.5 - 2m. We conduct four experiments by placing the radar at 2,4,6,8m distances from the closest tag. For each experiment, we collect 100 frames with a single tag switched on, two tags switched on, and all three tags switched on. Finally, we calculate the error in localization of the tags as the number of active tags increases in the environment. The results are provided in figure~\ref{fig:multi_tag_localization}. We see that the error in localization increases from 0.25 to 0.28 when three tags are active but still remains comparable to the single tag performance. The Code based modulation scheme in R-fiducial quite effectively separates the signal from each tag and provides reliable and accurate localization, even in the presence of multiple tags. }


\subsection{\name on the field}
\textbf{Mobility Experiments:} 
To show the performance of \name in more realistic settings we perform the experiments by mounting the radar on a car moving at varying speeds to evaluate the detection performance of \name. Figure \ref{fig:mobility_sim}b shows that \name maintains its high detection rate even under mobility scenarios. To further test our system at higher dopplers, we simulate speeds upto 150kmph and plot the distance estimation error in figure \ref{fig:mobility_sim}a. The plot shows that \name can detect the tag even in case of high doppler and provide accurate distance estimation. \name's spread spectrum codes provide resiliency to doppler by design (section~\ref{sec:section4.1}). Figure~\ref{fig:mobility_sim}c shows the deployment of radar on the car for mobility experiments.

  \begin{minipage}{\textwidth}
  \begin{minipage}[b]{0.49\textwidth}
    \centering
    \includegraphics[width=0.9\linewidth]{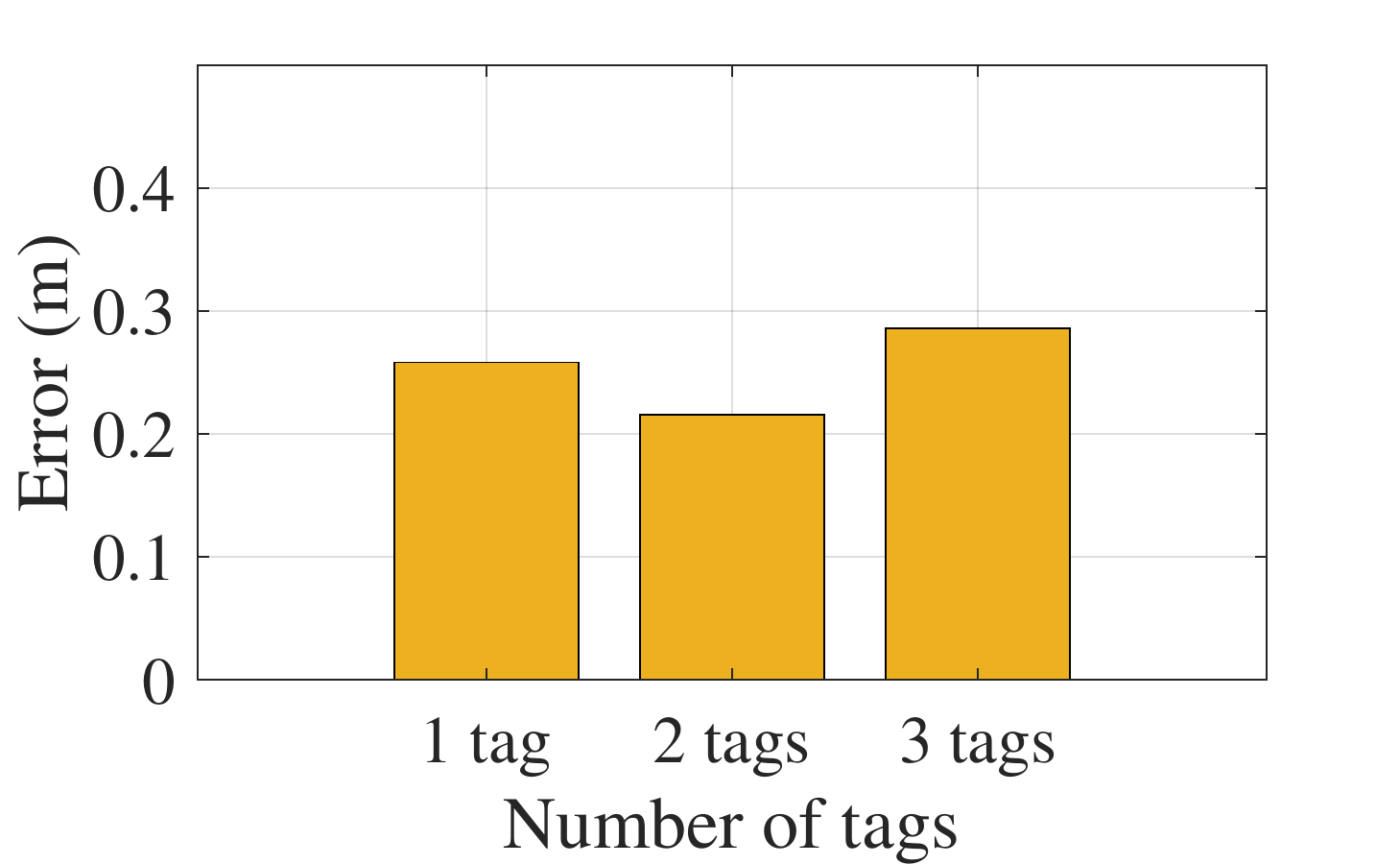}
    \captionof{figure}{Localization error for multi-tag}
    \label{fig:multi_tag_localization}
  \end{minipage}
  \hfill
  \begin{minipage}[b]{0.49\textwidth}
    \centering
    \begin{tabular}{|c|ccl|}
\hline
                  & \multicolumn{3}{c|}{\textbf{AUC}}                               \\ \hline
\multicolumn{1}{|l|}{} & \multicolumn{1}{l|}{\textbf{2.4 ms}} & \multicolumn{1}{l|}{\textbf{9.6 ms}} & \textbf{38.4 ms} \\ \hline
\textbf{Config 1} & \multicolumn{1}{c|}{0.985} & \multicolumn{1}{c|}{0.998} & 0.999 \\ \hline
\textbf{Config 2} & \multicolumn{1}{c|}{0.887} & \multicolumn{1}{c|}{0.986} & 0.999 \\ \hline
\textbf{Config 3} & \multicolumn{1}{c|}{0.627} & \multicolumn{1}{c|}{0.937} & 0.999 \\ \hline
\end{tabular}%
      \vspace{1cm}
      \captionof{table}{Area under curve for multi-tag experiment}
      \label{tab:multi_tag_auc}
    \end{minipage}
    \vspace{0.5cm}
  \end{minipage}

\begin{figure}[t!]
    \centering
    \includegraphics[width=\linewidth]{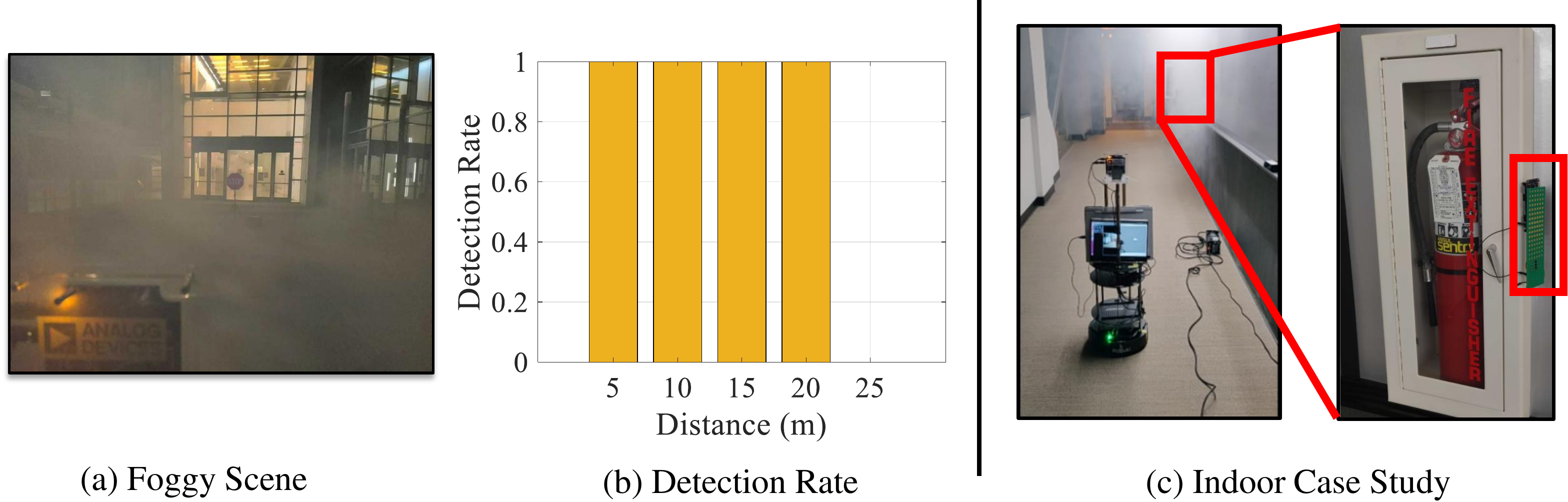}
    \caption{{\name on the field. (a) We conducted experiments in foggy weather outdoors to show the benefits of using \name in bad weather conditions. (b) shows the reliabel detection rate of \name even when the visibility is low. (c) We also conduct an indoor case study of fire extinguisher detection during smoke showing that \name can enable a myriad of applications both indoors and outdoors.}}
    \label{fig:badweather}
\end{figure}

\textbf{Bad Weather Experiments}
One of the main features of \name is to enable detection of traffic infrastructure in adverse weather. We perform experiments to showcase the reliability of \name during low visibility conditions (fog) where a camera or LiDAR based system would fail. Figure \ref{fig:badweather} shows the performance of \name in case of fog. \name maintains its high reliability even in adverse weather conditions which is vital for both indoor and outdoor applications.

We also perform a case study in an indoor scenario. Specifically, we place a tag in a corridor to mark a fire extinguisher. We generate smoke to emulate fire using a smoke machine. We simultaneously collect data from the camera and the radar. We show that with the presence of \name, radar is reliably locate the tag in an environment where the camera fails to see anything due to heavy smoke. Figure~\ref{fig:badweather}c shows the experiment snapshot. A video demonstration is provided in the supplementary material and at \textit{https://streamable.com/7ax59s}.


\subsection{Microbenchmark: Reliability of Code based modulation}

\begin{figure}[t!]
    \centering   
    \includegraphics[width=\linewidth]{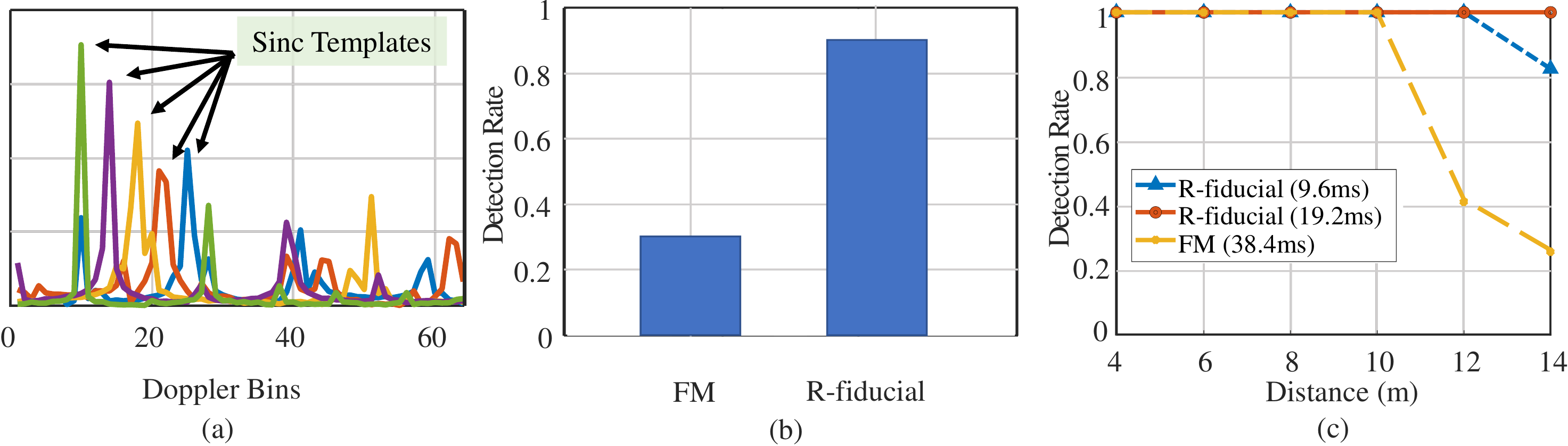}    
    \caption{Comparison of a frequency modulation based (FM) scheme\cite{soltanaghaei2021millimetro} vs \name code based scheme. (a) Shows the sinc templates for matching in FM method. All the template are different. (b) shows the detection rate in case of a moving car interference. (c) shows the detection rate trend with the distance (the observation time is given in brackets). \name outperforms the FM scheme, providing more reliable results in lesser latency.}  
    \label{fig:detection_comp}
\end{figure}

\textcolor{\hlcolor}{In this section, we evaluate the performance of our in-chirp code-based modulation (CDM) scheme design against a frequency-based modulation (FM) baseline. We use Millimetro\cite{soltanaghaei2021millimetro} as the baseline implementation of the FM scheme. For the FM method, we choose five different modulation frequencies between 200-600 Hz, which are in the same range as used in \cite{soltanaghaei2021millimetro}. Since FM is an inter-chirp modulation scheme, the templates for matching are generated in the doppler domain (section~\ref{sec:mod_techniques}). Figure \ref{fig:detection_comp}a shows the different sinc templates generated for these frequencies in the doppler domain. The total available bandwidth for sinc templates is limited by the total doppler bandwidth available, which depends on the time between two consecutive chirps (section~\ref{sec:radar_primer}). \cite{soltanaghaei2021millimetro} uses a chirp duration of 1ms, which provides a maximum doppler of 1kHz. To compare the detection performance of the two schemes, we perform two experiments:}

\textbf{a) Reliability in presence of clutter}:\textcolor{\hlcolor}{ In this experiment, we simulate the tag backscattering system in Matlab and generate the samples received at the radar for our CDM and the FM scheme. The radar used in our implementation is the same as that in \cite{soltanaghaei2021millimetro}. The duration of the chirp is set to 600us, which provides a better maximum detectable Doppler frequency of 1.67 kHz. We focus on particularly challenging scenarios, for example, when a vehicle is moving close to the tag in an outdoor setting, creating a much stronger peak in the doppler dimension due to the larger area and metallic body of the vehicle. We consider different speeds varying from 0-20 km/h. Figure \ref{fig:detection_comp}b shows the detection rate of the two systems. \name's CDM scheme outperforms the detection rate of the FM scheme as it does not get affected by clutter from the vehicle. In general, a typical car moving on a highway at more than 100mph can generate a doppler up to 5 kHz, which is already larger than the available Doppler bandwidth in the current system and radar. For achieving reliability from doppler clutter, the FM scheme needs to use a radar that allows smaller chirp duration. For example, in certain long-range radars with smaller chirp duration (for, e.g., 30us in TI LRR radar~\cite{dham2017programming}), the doppler bandwidth could be higher (up to 25 kHz), allowing for switching frequencies higher than 5kHz. Here, it may appear that the baseline approach only fails for the first 25\% (5kHz out of 25 kHz) of the time and works otherwise. However, the inter-chirp modulation FM scheme creates the sinc pattern in the entire doppler bandwidth, meaning the template has side lobes present in the entire bandwidth. Even if one chooses a very high switching frequency, the template matching can still trigger a false positive as one of the side lobes could overlap with the doppler from an object. This problem significantly lowers the number of tags that can be reliably supported. This is actually similar to the reason why FDMA has less capacity and supports less number of users than CDMA~\cite{cheng1992total}. \name employs the latter approach, which is more flexible and truly independent of the doppler interference from objects. This increases the scope of applications to a great extent in both indoor and outdoor scenarios.}


\textbf{b) Distance measurement:} \textcolor{\hlcolor}{In this experiment, we implement the baseline FM scheme~\cite{soltanaghaei2021millimetro} on our tag hardware to make a fair comparison between the two schemes. We compare the detection rate of the tag using two modulation schemes at different distances from the tag in an indoor setting. We move the tag from distances starting from 2m to 14m. We also evaluate the time taken for the measurement. Figure~\ref{fig:detection_comp}c shows the performance of the two modulation schemes. The detection rate of the FM scheme starts to fall after 10m, while \name's code-based scheme continues to provide reliable detection. Note how \name provides more reliable detections in much lower latency compared to the FM system. Please note that in ~\cite{soltanaghaei2021millimetro}, the authors create two different types of tags. The smaller tags are used to estimate the power consumption, while the larger tags are used for distance measurement and on-field testing. The larger tag contains high gain antenna arrays~\footnote{A high gain narrow beamwidth antenna similar to \href{https://www.ebay.com/itm/113705631458}{this (hyperlink)} is used in ~\cite{soltanaghaei2021millimetro}}. These antenna arrays provide a high gain but at the expense of very narrow azimuthal beamwidth (12 deg), which is impractical for many applications involving a radar fiducial. Consequently, in ~\cite{soltanaghaei2021millimetro}, the evaluations are performed on normal incidence where the antenna gain is very high to get a longer range. The retro-reflectivity is not evaluated because a big setback of using narrow beamwidth antennas is that they can not work arbitrary incident angles except for normal incidence. On comparing the \name and the FM scheme on similar hardware, we find that \name’s CDM scheme provides higher reliability (experiment b). It also keeps the latency of detection low and allows the capability of getting processing gain by using longer observation times.}

\textcolor{\hlcolor}{Another important factor affecting the reliability of detection is optimality in template matching. In a frequency-based design (FM), detection is performed by matching the templates which are formed from the sinc patterns. However, the asynchronous operation of the reader and tag can cause the phase of sinc peak to take any random value (refer equation 2 in ~\cite{soltanaghaei2021millimetro}). Because of this unknown random phase, while performing template matching, we are forced to match the absolute values of sinc pattern. In contrast, CDM scheme does correlation in time domain where we even match the phase values of the signal. The phase level matching provides the optimal SNR for detection which is always better than amplitude based absolute matching in FM schemes, similar to the optimality of co-phasing\cite{MRC_optimality} in maximal ratio combining. This results in lesser reliability when doppler clutter is present (experiment a) or when the signal becomes weaker at longer distances (experiment b).}


\section{Related Work}
\label{sec:related}

\name presents the optimization framework, design, implementation of mm-wave backscatter tags, which are low-power and can be attached to existing traffic infrastructure. \name tags can be read using a single chirp and can be read by most COTS automotive radars, including beam scanning or digital beamforming radars. \name code-design allows it to be doppler-resistant, scalable to multiple tags and resilient to the environmental reflections, multi-path and interference from other tags. 

\noindent \textbf{Van Atta Array based mmwave tags:} Use of Van Atta array concept to design retro-reflective tags has been explored in several past works from early 2007. In  ~\cite{miao2017passive, ren2007new, ang2018passive}, retro-reflective antenna array designs for microwave bands have been considered. Vitaz et.al\cite{vitaz2011enhanced,vitaz2010tracking} designed a Van Atta based tag for 26GHz (mmwave) to be used for object tracking and identification. However, it uses power-hungry PIN diode switches for data modulation and are read by a pulse radar -- which is not popularly used for automotive applications. REITS \cite{REITS} perform simulations based on using millimeter wave tags for transportation systems. We contrast with REITS in two aspects: Firstly, we use PN code sequences compared to the On-Off keying used by REITS, to modulate the backscatter signal. Associating a unique PN code to each tag offers us resilience against moving vehicles. Secondly, our method allows us to detect and localize tags within a single chirp duration. In Millimetro\cite{soltanaghaei2021millimetro}, a promising design of retro-reflective tag at 24 GHz is presented in which a long detection range has been shown by using high gain antennas. We on the other hand, design a stand-alone tag with printed antennas and provide a framework to choose the number of antennas in order to support diverse applications. Their design uses frequency modulation-based techniques and matched filtering with sinc templates becomes unreliable in case of doppler shift due to tag/vehicle movement (Section \ref{sec:evaluation}). In contrast \name's design makes it completely robust against the doppler speeds one would encounter in automotive radar applications. Furthermore, Millimetro needs an entire radar frame that contains several chirps, in order for the detection to take place. This is not suitable for beam scanning radars that send individual chirps in spatially different directions.
\textcolor{\hlcolor}{Hester et.al\cite{Hester_longrange_tag} present an integrated ammonia wireless sensor with Van atta array structure. This work also uses frequency modulation and suffers from unreliable detection in situations where there is a lot of moving background clutter in the measurements}. Another line of works \cite{nolan2021ros, trzebiatowski202060} design cross-polarized van-atta array-based tags at higher mmwave frequencies (60/77 GHz). The design of \cite{nolan2021ros} encodes the data in RCS patterns of tags which requires radar reading  the tag to support multiple polarization, thereby highly limiting to usable by specialized radars only. \name can provide robust detection performance for realistic driving scenarios with any COTS automotive radars. Another work mmTag\cite{mazaheri2020millimeter, mazaheri2021mmtag} designs millimeter wave backscatter tags to enable low-power high-throughput wireless links. However, their target application is communication and not sensing infrastructure.

\textbf{3D printed backscatter tags:} The advancements in 3D printing have become a low-cost alternative to rapid manufacturing of backscatter tags. Recent works \cite{hester2016inkjet,inkjet_print_antenna,90GHz_printed_antenna,inkjet_Eid} demonstrate a printable backscatter Van Atta arrays that work at mmwave frequencies. The antennas on the tag can be manufactured by simply using a conductive ink on commercially available Kapton polyimide film\cite{KAPTON} to act as a dielectric. With these advancements, our design can be 3D printed to enable rapid development of \name tags with different antenna designs as per the requirements explained in section 3.


\textbf{Chipless RFID:}
Hester et.al \cite{inkjet_print_antenna,hester2016inkjet} present a chipless RFIDs using Van Atta Array where they identify the tags based on the tag's resonant frequencies. A common approach is to use notch frequencies to create an identifying signal from RFIDs \cite{balbin2009novel, gao2018livetag, jalaly2005capacitively}. RoS \cite{nolan2021ros} proposes changing polarization to identify backscatter tags. However, most of the automotive-grade radars  \cite{ramasubramanian2017awr1243,meyer2019automotive} used in modern vehicles are equipped with arrays of linearly polarized patch antenna arrays, which do not allow a polarimetric modulation approach to work. In other words, a polarization-based design would involve significant hardware changes in the current automotive radars, which defeats the goal of non-intrusive traffic infrastructure augmentation. 

\textbf{MIMO backscatter:} Conventional RFID based architectures~\cite{rao2006theory, preradovic2009design} can identify themselves well to the reader, but do not scale well to longer distances because RFID tags use a single antenna thus limiting the radar crosssection(RCS) of those tags. In \cite{malecki2013rfid, garcia2018passive} RFIDs are used for traffic sign detection but require extra hardware to be placed on vehicles. Tunnel scatter\cite{varshney2019tunnelscatter} uses tunnel diodes that have negative resistance to terminate the tag antenna. Negative resistance makes it possible to realize a reflection coefficient of more than 1 and hence increasing the backscatter signal power at the expense of consuming power in the tunnel diode. \cite{gong2018backscatter} makes use of polarization diversity to propose a tag design to steer the backscatter signal. To further enhance the RCS , \cite{bidirec_amplifier} chooses an active backscatter approach and introduce bidirectional amplifiers in the array. However, these works include an active element in their design and are not suitable for low power applications. 

\begin{figure}[t!]
    \centering
    \includegraphics[width=0.8\linewidth]{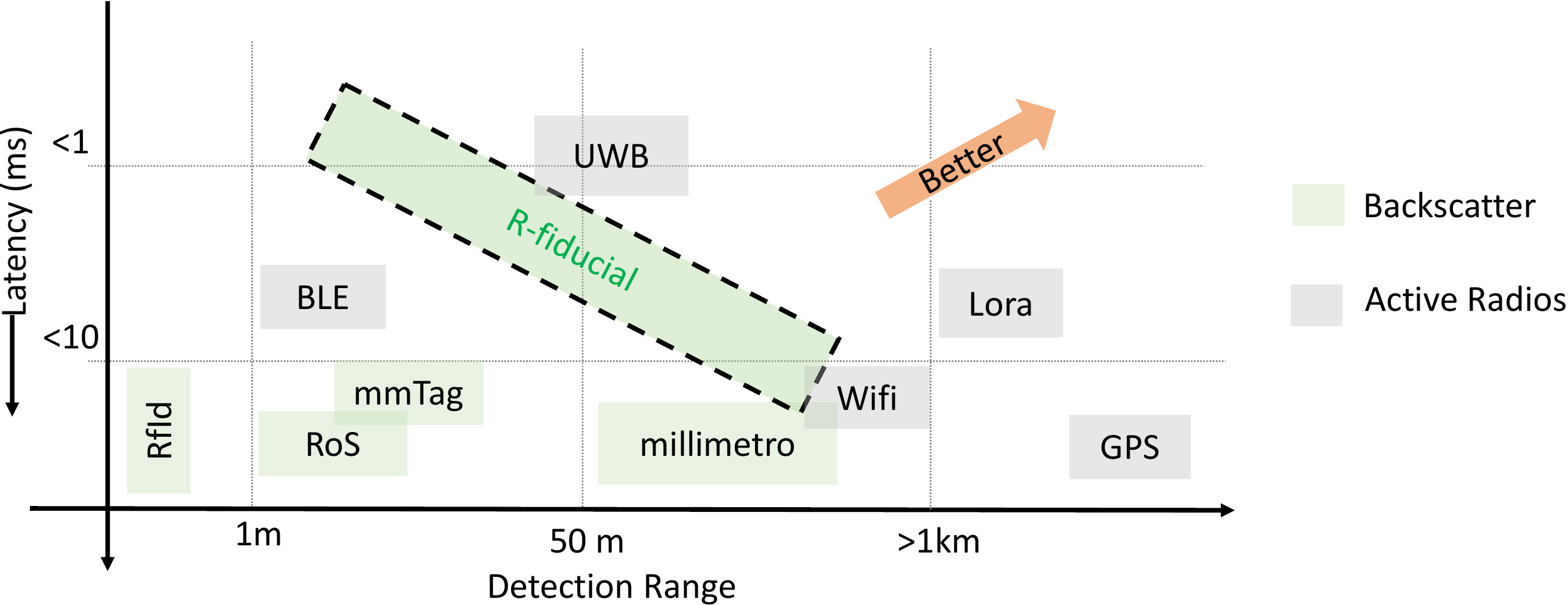}
    \caption{{Comparison of different sensing technologies in terms of latency vs detection range. \name provides a flexibility in operation and hence can be used widely in indoor and outdoor applications.}
    }
    \label{fig:comparison}
\end{figure}

We summarize the features of mmwave backscatter technologies and compare them with \name in table~\ref{tab:comparison}. We further compare \name with the existing sensing technologies in terms of latency and range of operation in figure~\ref{fig:comparison}. \name provides flexibility in its operation which truly makes it a technology that can be ubiquitously deployed for many indoor and outdoor applications.


\begin{table*}[]
\centering
\resizebox{\textwidth}{!}{%
\begin{tabular}{c|c|c|c|c}
 &
  \textbf{Radar Type} &
  \textbf{Range of operation (m)} &
  \textbf{Sensing latency (ms)} &
 
  \textbf{Doppler performance} \\ \hline
\textbf{RoS~\cite{nolan2021ros}}        & Custom & \textless{}25    & Depends on car speed         & Precise self-tracking required \\ \hline
\textbf{Millimetro~\cite{soltanaghaei2021millimetro}} & Digital BF only    & \textgreater{}25 & 64   & Doppler correction required    \\ \hline
\textbf{\name} &
  Analog/Digital BF &
  \textbf{6 - 25+} &
  \textbf{$\approx$0.5 - 38} &
  \textbf{Robust against doppler} \\ \hline
\end{tabular}%
}
\vspace{15pt}
\caption{Comparison of \name with the past approaches based on different design requirements, BF: Beamforming}
\label{tab:comparison}
\end{table*}

\section{Discussion \& Limitations:}
\textbf{Non Line of Sight(NLoS) operation of \name tags:} \textcolor{\hlcolor}{\name tags do not require the tags to be in line of sight with radar. As long as the signal from the tag reaches back to the radar, \name tags can be detected. The proposed tag detection algorithm do not constrain the tags to be in line of sight. In contrast to cameras/lidars, radars can still work in conditions where NLOS occurs due to occlusions and some blockages such as cars/trees~\cite{palffy2019occlusion}. Moreover, the detection algorithm would work even when the tag's reflection reaches the radar upon bouncing off different objects. However, a Line of Sight(LoS) operation achieves a longer range of detection than NLoS conditions due to the signal power degradation in NLoS.}

\textbf{Production and maintenance of \name tags:} \textcolor{\hlcolor}{ \name tags are manufactured on a 15 cm $\times$ 7 cm printed circuit board (PCB) with RF switches soldered on the PCB. When produced at large scale, it costs around \$5 per \name tag PCB. In our prototype, we use a low-power Microcontroller(MCU) to turn the RF switch ON and OFF. For practical deployment, an application-specific integrated circuit ASIC containing the RF switches and the control logic can be manufactured to keep the cost low. \name tags are battery powered and are estimated to have a life-time of 1-2 years when powered by a coin cell. Frequent replacement of these batteries can be avoided by using energy harvesting techniques like mini-solar cells and these tags can be deployed outdoors sustainably with very low maintenance cost.}

\textbf{V2V communications:} \textcolor{\hlcolor}{\name tags are passive and retro-reflect the signals without using any active amplification at the tag. In applications where the power consumption is not a concern,  (For example when the tags are used on vehicles, they can be powered by the vehicle's battery), active devices like bi-directional amplifiers can be introduced into the Van Atta array structure to boost the reflected signal power. This modification would help \name tags to enable reliable Vehicle to Vehicle(V2V) communications that will advance the capabilities of Autonomous driving assistance systems.}

\textbf{2D Retro-reflective tags:} \textcolor{\hlcolor}{\name tags uses a series fed patch antenna to increase the gain and consequently have finite vertical field of view. So, the tag can good retroreflectivity when the radar and the tag are at a similar elevation from the ground. This limits the deployability of tags at arbitrary elevations. This issue can be resolved by developing a 2D van atta array that can retro-reflect signals from any incident direction in the horizontal or vertical plane. Some early approaches\cite{2d_van_atta,star_van_atta} achieve 2D retro-reflectivity by arranging patch antennas in a 2D arrangement but they do not modulate the reflected signal. Our future work is to integrate our modulation technique with these 2D Van-atta arrays.}

\section{CONCLUSIONS}
\noindent \name provides a scalable solution to reliably augment the mmwave sensing infrastructure that is compatible with commercial mmwave radar technology to enable several indoor and outdoor use cases. \name uses Van-atta array design to enable retro-directivity so that a reader radar can read it from a wide range of angles and provides a framework to design a suitable van-atta architecture. Secondly, \name distinguishes itself from other objects encountered on roads by creating a unique signature in form of switching frequency. Finally, each \name is encoded using a specific code, that creates a specific identity for each tag (eg, stop sign, traffic light). Using a code based identification provides scalability required to support multiple tags working simultaneously. With all these features, \name can be reliably used in applications where visual sensors do not work because of poor visibility.

\bibliographystyle{ACM-Reference-Format}
\bibliography{main}

\end{document}